\documentstyle[12pt,epsfig]{article}

\newcommand{\mathbb}{{\sf \hspace*{-0.96ex}}}

\title{Nonlocal constitutive laws generated by matrix functions: Lattice Dynamics Models and their Continuum Limits}

\author{{\bf Thomas M. Michelitsch\footnote{Corresponding author, e-mail~: michel@lmm.jussieu.fr, www~: http://bit.ly/champs-fractelysees}\,,\,\, Bernard Collet} \\ \\
Institut Jean le Rond d'Alembert, CNRS UMR 7190 \\
Universit\'{e} Pierre et Marie Curie, Paris 6\\
4 Place Jussieu\\
75252 Paris cedex 05\\
France
\\ \\ \\
 {\bf Xingjun Wang}\\ \\
Department of Physics\\
Qiannan Normal College for Nationalities\\
LongShan Road\\
558000 Duyun, Guizhou \\
P.R. China
\\ \\ \\
{\it Submitted Manuscript}
\\ \\ \\
{\small \it \jobname .tex }
}
\begin{document}
\maketitle

\begin{abstract}
We analyze one-dimensional discrete and quasi-continuous linear chains of $N>>1$ equidistant and identical mass points with   periodic boundary conditions and generalized nonlocal interparticle interactions in the harmonic approximation. We introduce elastic potentials which define by Hamilton's principle discrete ``Laplacian operators'' (``Laplacian matrices'') which are operator functions ($N\times N$-matrix functions) of the Laplacian of the Born-von-Karman linear chain with next neighbor interactions. The non-locality of the constitutive law of the present model is a natural consequence of the {\it non-diagonality} of these Laplacian matrix functions in the $N$ dimensional vector space of particle displacement fields where the periodic boundary conditions (cyclic boundary conditions) and as a consequence the (Bloch-) eigenvectors of the linear chain are maintained. In the quasi-continuum limit (long-wave limit) the Laplacian matrices yield ``Laplacian convolution kernels'' (and the related elastic modulus kernels) of the non-local constitutive law. The elastic stability is guaranteed by the positiveness of the elastic potentials. We establish criteria for ``weak" and ``strong" nonlocality of the constitutive behavior which can be controlled by scaling behavior of material constants in the continuum limit when the interparticle spacing $h\rightarrow 0$.
The approach provides a general method to generate physically admissible (elastically stable) {\it non-local constitutive laws} by means of ``simple'' Laplacian matrix functions. The model can be generalized to model non-locality in $n=2,3,..$ dimensions of the physical space.
\end{abstract}

\section{Introduction}
There is a wide range of materials and deformable structures at nanoscale for which local constitutive descriptions appear to be
inappropriate. For instance material systems such as carbon nanotubes show on the nanoscale size effects
which strongly suggest the importance of non-local inter-particle interactions and as a consequence the
non-locality of the constitutive laws which cannot be covered by classical local continuum theories.
The importance of non-local constitutive behavior was raised by several authors already in the sixties of the last century such as Eringen (1972, 1983, 1992, 2002), Kr\"oner (1966), Krumhansl (1968) and  Kunin (1982). Born-von-Karman models and their continuum counterparts account only for local inter-particle interactions represented by next- or close neighbor particle springs (Askar, 1985; Born \& Huang, 1954; Scrivastava, 1990) and as a consequence
these models predict dispersion free wave propagation in the continuum limit (long-wave limit). Hence
these classical models are only able to describe the effects of local inter-particle interactions on the
vibrational properties.

In order to account for size effects in elastic nano-structures, theoretical models with nonlocal interactions were raised recently for beams, plates and shells (Challamel \& Wang 2008; Lu et al., 2007; Lim, 2010; Reddy, 2007; and references therein).

On the other hand there are numerous references which approached the subject of non-locality for crystalline materials in the framework of lattice dynamics by accounting for interactions which include first second, and close neighbor particles of multiple order where the interparticle interactions as a rule are modeled by linear interparticle springs (Born \& Huang 1954; Maradudin et al., 1963; Remoissenet \& Flytzanis, 1984; Cadet, 1987). However, there are not many {\it non-local models} which link the continuum approach and the lattice dynamics approach (e.g. Askar, 1982, 1985; Eringen \& Kim, 1977; Rosenau, 2003).
A further weak nonlocal lattice model and its continuum limit was presented by Pouget (2005) in the context of a non-linear lattice model to describe the complex dynamics of pattern formation and by Collet (1993) in the framework of model of solitonic wave propagation in cubic cristals. As a rule in those references the harmonic approximation to non-locality was
introduced as simple nonlocal linear springs where the nonlocal spring constants define a nonlocal convolutional operator
in the continuum limit. Despite those advanced models there is further need of analytical models to cover aspects
of non-locality and interconnecting both the lattice dynamics approach and the continuum approach.

To this end the goal of the present paper is to introduce simple non-local periodic 1D lattice (linear chain) models which lead to non-local behavior in the discrete
linear chain, and to deduce the corresponding non-local constitutive law in the continuum limit. We show in the present paper that scaling relations of the material constants in the continuum limit are required in order to guarantee that the elastic energy remains finite. The starting point for our models are positive elastic energies which define by Hamilton's variational principle generalized Laplacian operators with all required ``good properties'' of Laplacians including elastic stability.

So far there is a huge lack of non-local constitutive models in the literature accounting for the interlink between lattice dynamics and continuum approach. In most classical
works on non-locality such as of Eringen (1983, 2002), Lazar et al. (2006), Maugin (1979, 1999) and others such as Kr\"oner (1967), Kunin (1982), Eringen (1972, 1983, 1992) non-locality is {\it phenomenologically introduced} for some
constitutive convolution kernels of simple forms or by inclusion of some ``convenient'' higher order gradient type models such as by Eringen (2002), Lazar et al. (2006), and Maugin (1979). However, most of those models are pure continuum models and were not linked to lattice dynamics models.

A three-dimensional lattice model was introduced by Eringen and Kim (1977) which analyzed the link between lattice dynamics and non-local elasticity in three dimensions by considering the continuum limit of small interparticle distance for non-local harmonic springs to describe harmonic far-range interactions. Nevertheless, there seems to be still a lack of more general approaches which deduce non-local constitutive continuum models rigorously from discrete lattice dynamics models. The present paper aims to contribute in this respect by introducing non-locality by a discrete (linear chain) lattice approach and by analyzing its continuum limit. The starting point are harmonic elastic potentials defined on a periodic linear chain which include non-local
harmonic interparticle interactions and define by Hamilton's variational principle discrete non-local constitutive laws
of matrix forms which take in the continuum limit non-local convolutional forms.

The paper is organized as follows:
We introduce a one-dimensional linear chain model with non-local harmonic interactions and periodic boundary conditions. We generate the non-local constitutive behavior by constructing elastic potentials which lead via Hamilton's variational principle to ``Laplacian operators" which are operator functions (matrix functions in the $N$-dimensional space of particle displacements) of the local ``Laplacian" of the next-neighbor Born-von Karman linear chain model. In a sense we conceive the ``local Laplacian'' operator as ``{\it generator}" of ``non-local Laplacian'' operators.

We analyze the vibrational dispersion relation (negative eigenvalues of the Laplacian) of the
discrete chain with periodic boundary conditions and analyze its continuum limit (long-wave limit) rigorously. The Laplacian
then takes in the long-wave limit the form of the non-local convolutional constitutive law which contains also the full information on the elastic modulus kernel. The ``degree of nonlocality" of these kernels is sensitive on the scaling behavior of the material constants in the continuum limit.




\section{Discrete non-local model for the quasi-continuous 1D linear chain}
We consider a linear chain of $N$ identical particles where $N>>1$ is assumed to be ``large" and $u_p=u(x_p)$ denotes a field variable
such as the displacement field associated to particle $p$ located at lattice points $x_p=ph$ ($p=0,..,N-1$). We assume equidistant particles with interparticle spacing $h$ (lattice constant) and identical particle masses $\mu$.
The length of the linear chain is indicated by $L=Nh$. We use a compact notation by means of distributions and functional calculus.
We assume the periodic boundary conditions (cyclically closed chains)

\begin{equation}
\label{periodic}
 u(x_p)=u(x_{p+N})
\end{equation}
Due to the periodicity of the boundary conditions we use throughout this paper particle indices $p$ cyclically, such that $0\leq p \leq N-1$ (When we generate matrix functions with indices $p \notin \{ 0,1,..,N-1\}$ outside this set, the cyclic index convention maps them back into this set by replacing $p \rightarrow p \,\,\, mod(N) \in \{ 0,1,..,N-1\}$).
We introduce the following notation: Summations over the discrete lattice points are written has
\begin{equation}
 \sum_{p=0}^{N-1} f(x_p) = \int_{-\infty}^{\infty} \rho(x)f(x){\rm d}x,
\end{equation}
where we introduce the particle number density distribution function $\rho(x)$ as
\begin{equation}
 \label{distfu}
\rho(x)= \sum_{p=0}^{N-1}\delta(x-x_p),
\end{equation}
where $\delta(\xi)$ indicates Dirac's $\delta$-function. $\rho(x)$ indicates the number of particles per unit-length and is normalized as
\begin{equation}
 \label{norm}
\int_{0}^{L} \rho(x){\rm d}x = N.
\end{equation}

The total energy (Hamilton function) of the discrete linear chain can be written as a functional
\begin{equation}
\label{hamden}
 H= \int_{-\infty}^{\infty} \rho(x){\cal H}(x){\rm d}x.
\end{equation}
Where ${\cal H}$ denotes a Hamiltonian density (energy associated with mass points $x=x_p$).
Note, due its definition this Hamiltonian ``density'' has the physical dimension of an energy.
The Hamiltonian density can be written as

\begin{equation}
\label{massinteg}
{\cal H}(x)=  \frac{\mu}{2}\left(\frac{\partial u}{\partial t}\right)^2
+ {\cal V}(x,h),
\end{equation}
where ${\cal V}(x,h)$ indicates the elastic energy of {\it non-local harmonic interactions} of a particle located at $x=x_p$ with the other particles of the chain. First let us consider a harmonic nonlocal interaction which is governed by
$m$-order differences of the displacement field which we introduce as follows

\begin{equation}
\label{mintroduce}
{\cal V}_m(x,h) = \frac{\mu\Omega_m^2}{2}\left[\left(D(h)-1\right)^mu(x)\right]^2,
\end{equation}
with the dimensional constant $\Omega_m$ having physical dimension of a frequency.
In (\ref{mintroduce}) we further introduced the shift operator which shifts the argument $x$ of the displacement field $u(x)$
by the shift $h$, namely\footnote{The shift operator has all Abelian group properties of the shift operation itself, such as $ D(0)=1$, $D(h_1)D(h_2) = D(h_1+h_2)$, $D^s(h)=D(sh)$, and $D^{-1}(h)=D(-h)$..}

\begin{equation}
 \label{shiftdef}
D(h)u(x)=u(x+h) ,\hspace{1cm} D(h)u_p=u_{p+1} ,\hspace{0.5cm} u_p=u(x_p)
\end{equation}
(where ${\rm p=0,1,..N-1,\,cyclically}$).

Despite we focus in this paper on integer orders $m=1,2,..\in \mathbb{N}$, in general the model is well defined for any any {\it positive real number $m>0\in\mathbb{R}$} where in the non-integer case fractional calculus comes into play. Although the non-integer case is not subject of the present paper, we will come back to this important point in brief subsequently as the fractional case of non-integer $m$ is an important source of non-locality.

Subsequently it will be convenient to utilize the following representation of the shift operator which holds for sufficiently smooth (infinitely often differentiable) fields $u(x)$ as
\begin{equation}
 \label{shift}
D(h)=\exp{(h\frac{d}{dx})}= \sum_{m=0}^{\infty}\frac{h^m}{m!}\frac{d^m}{dx^m},
\end{equation}
acting on $u$ by generating the Taylor expansion of $u(x+h)$ around $x$ with respect to shift $h$.
The long-wave limit (continuum approximation) guarantees that the field $u(x)$ is sufficiently smooth that the shift operator can be represented by (\ref{shift}).
We can hence write the difference of first-order, and second-order differences ($h$ denotes the distance of next-neighbor particles) of displacement fields

\begin{equation}
\label{nextneigh}
\begin{array}{l}
\displaystyle \  u(x_{p+1})-u(x_p)= (D(h)-1)u(x)|_{x=x_p},\nonumber \\ \nonumber \\
 \displaystyle \ u(x_{p+2})-2u_(x_{p+1})+u(x_p)= (D(h)-1)^2u(x)|_{x=x_p},
 \end{array}
\end{equation}
and so the definition of the $m^{th}$ order difference in the elastic energy (\ref{mintroduce}) is given by

\begin{equation}
 \label{evalop}
\left(D(h)-1\right)^mu(x)|_{x=x_p} = \sum_{s=0}^m\frac{m!}{s!(m-s)!}(-1)^{m-s}u(x_p+sh).
\end{equation} 

We observe in (\ref{evalop}) that the harmonic interaction (\ref{mintroduce}) of a particle located at $x=x_p$ involves interaction of a particle $p$ with its $m$ closest neighbor particles located at $u(x_p+sh)$ ($s=1,..,m$) in the $(+)$-direction.
Note that the case $m=1$ reproduces the (local) next-neighbor (Born-von-Karman) linear chain model.
The summation over all particles $p$ in the total elastic energy contained in (\ref{hamden})
guarantees that both, the $m$ closest neighbor particles in forward (+) and in $(-)$- directions symmetrically are taken into account. Hence (\ref{hamden}) is equivalent to the symmetrized form of the elastic potential $\frac{{\cal V}_m(x,h)+{\cal V}_m(x,-h)}{2}$.

We can now construct generalized harmonic non-local elastic potential by superposing (\ref{mintroduce}) involving all orders of $m$ and hence a great number of neighbor particles involved. We can write such potentials in symmetric form as

\begin{equation}
 \label{neighborsfunction}
{\cal V}_f(x,h)= \frac{1}{2}\sum_{m=1}^{\infty}a_m\left({\cal V}_m(x,h)+{\cal V}_m(x,-h)\right),
\end{equation}
where the cyclic index convention guarantees that (\ref{neighborsfunction}) is well defined for any order $m$.
Then the total elastic energy of (\ref{hamden}) writes explicitly

\begin{equation}
\label{potfunctionalf}
\begin{array}{l}
\displaystyle \ V_f[u] = \frac{\mu}{2}\sum_{m=1}^{\infty}\Omega_m^2\int_{-\infty}^{\infty}{\rm d}x\rho(x)\left\{\left(D(h)-1\right)^m u(x)\right\}^2 \nonumber \\ 
\\
 \displaystyle \ =\frac{\mu}{4}\sum_{m=1}^{\infty}a_m\Omega_m^2\int_{-\infty}^{\infty}{\rm d}x\rho(x)[\left\{\left(D(h)-1\right)^mu(x)\right\}^2\nonumber \\ 
 \\
 \displaystyle \ +\left\{\left(D_{x}(-h)-1\right)^m u(x)\right\}^2].
 \end{array}
\end{equation}

For our convenience we introduce the (positive) ``characteristic function'' $f(\lambda)$ which has the same coefficients
$a_m\Omega_m^2$ as (\ref{neighborsfunction}) with (\ref{mintroduce}), namely
\begin{equation}
 \label{flam}
f(\lambda)= \sum_{m=1}^{\infty}a_m\Omega_m^2\lambda^m \geq 0.
\end{equation}
Without loss of generality we introduce the coefficients $a_m$ taking only the values $a_m=0,\pm 1$ and its absolute value is represented  either by $|a_m\Omega_m^2|=\Omega_m^2$ or zero if an order $m$ is absent. We will demonstrate subsequently that elastic stability requires that the entire series (\ref{flam}) is positive, i.e. $f(\lambda) >0$
for $0<\lambda<4$ and further we will show that translation invariance leads to $f(\lambda=0)=0$, i.e. $a_{0}=0$.
In the case that the series of the characteristic function $f$ breaks, i.e. $a_m=0, m>n$ we have only a kind of ``weak'' non-locality involving $n$ closest neighbor particle interactions.

The elastic potential (\ref{neighborsfunction}) contains the complete information on the non-local constitutive law. Now our goal is by applying Hamilton's variational principle to deduce the equation of motion
of the Hamiltonian system (\ref{hamden}). The variational principle defines in rigorous manner a  ``Laplacian'' operator which contains the full (in general non-local) constitutive information. If we conceive the fields
$u_p=u(x_p)$ as $N$ (Cartesian) components of a displacement vector ${\bf u}$, then the ``Laplacian" takes generally a non-diagonal $N\times N$ matrix representation where its non-diagonality in the $N$-dimensional displacement vector space describes the non-locality of the constitutive law.

In order to deduce this Laplacian we consider the total elastic energy term (\ref{mintroduce}) of order $m$ which can be rewritten (\ref{massinteg}) in the form

\begin{equation}
 \label{potenfu}
 \begin{array}{l}
\displaystyle V_m[u] = \frac{\mu\Omega_m^2}{2}\int_{-\infty}^{\infty}{\rm d\tau}\rho(\tau)\left\{(D_{\tau}(h)-1)^m u(\tau)\right\}^2 \nonumber \\ \nonumber \\
\displaystyle =\frac{\mu\Omega_m^2}{2}\sum_{p=0}^{N-1}\left\{(D_{\tau}(h)-1)^m u(x_p)\right\}^2,
\end{array}
\end{equation} 
where the particle distribution density $\rho$ of (\ref{distfu}) is taken into account.
The equation of motion is determined by Hamilton's principle in the form
\begin{equation}
 \label{lag}
 \begin{array}{l}
\displaystyle \mu\frac{\partial^2}{\partial t^2}u(x_p,t) = \Delta_{2m}(h)u(x_p,t), \nonumber \\ \nonumber \\
 \displaystyle \Delta_{2m}(h)u_p =- \frac{\partial V}{\partial u_p},
 \end{array}
\end{equation}
defining a ``Laplacian'' $\Delta_{2m}(h)$ which is to be determined
and $u_p=u(x_p,t)$ where $x_p$ denote the lattice points of the chain. Let us now evaluate
\begin{equation}
 \label{variation}
\frac{\partial }{\partial u_p}V_m[u] = \mu\Omega_m^2 \sum_{q=0}^{N-1}(D(h)-1)^mu_q \frac{\partial }{\partial u_p}(\{D(h)-1)^mu_q\}.
\end{equation}
To evaluate (\ref{variation}) we utilize $\frac{\partial u_{q+s}}{\partial u_p}=\delta_{p,q+s}=\delta_{p-s,q}$ so that
\begin{equation}
\label{consider}
\frac{\partial }{\partial u_p}(\{D_{\tau}(h)-1)^mu_q\} =  \sum_{s=0}^m\frac{m!}{s!(m-s)!}(-1)^{m-s}\delta_{(p-s)q},
\end{equation}
where we account for $D(sh)u_q=u(x_p+sh)=u_{p+s}$ shifts the particle index $q$ to $q+s$ and so we have
\begin{equation}
 \label{varrel}
\frac{\partial}{\partial u_p} D(sh)u_q = \frac{\partial}{\partial u_p} u_{q+s} = \delta_{p(q+s)}
= \delta_{(p-s),q} = \frac{\partial }{\partial u_{p-s}}u_q.
\end{equation}
Then we can further write for (\ref{variation})
\begin{equation}
 \label{var2}
\frac{\partial }{\partial u_p}V_m[u] = \mu\Omega_m^2 \sum_{q=0}^{N-1}(D(h)-1)^m  \sum_{s=0}^m\frac{m!}{s!(m-s)!}(-1)^{m-s} \delta_{(p-s)q} u_q,
\end{equation}
where we can identify the second binomial sum with
\begin{equation}
\label{ident}
\begin{array}{l}
\displaystyle\ \sum_{s=0}^m\frac{m!}{s!(m-s)!}(-1)^{m-s} \sum_{q=0}^{N-1} \delta_{(p-s)q}u_q= \nonumber \\ \nonumber \\
\displaystyle\ \sum_{s=0}^m\frac{m!}{s!(m-s)!}(-1)^{m-s}u_{p-s}=(D(-h)-1)^mu_p.
\end{array}
\end{equation}
So we arrive for (\ref{var2}) finally at

\begin{equation}
 \label{finally}
\begin{array}{l}
\displaystyle \frac{\partial }{\partial u_p}V_m[u] = \mu\Omega_m^2 (D(h)-1)^m(D(-h)-1)^mu_p  \nonumber \\ \nonumber \\
\displaystyle  \hspace{2cm} = \mu\Omega_m^2\left(2-D(-h)-D(h)\right)^mu_p.
\end{array}
\end{equation}
where $D(\pm h)u_p=u_{p\pm 1}$.
The equation of motion of the mass points $p$ is then determined by

\begin{equation}
 \label{Lag}
\mu\frac{\partial^2}{\partial t^2}u(x_p,t) = - \frac{\partial }{\partial u_p}V_m[u]
=\Delta_{2m}(h)u(x_p,t).
\end{equation}
The right-hand side of this generalized wave-equation represents the {\it constitutive law} and the operator acting on displacement field $u_p$
can be conceived as a {\it ``generalized Laplacian''}
\begin{equation}
\Delta_{2m}(h)u_p =  -\frac{\partial }{\partial u_p}V_m[u] ,\nonumber \\
\end{equation}
where this ``Laplacian''which corresponds to harmonic potential (\ref{potenfu}) of $m$ closest neighbor particle interaction
has the representation
\begin{equation}
\begin{array}{l}
 \label{lapla}
\displaystyle\ \Delta_{2m}(h) = (-1)^{m+1}\mu\Omega_m^2  (D(\frac{h}{2})-D(-\frac{h}{2}))^{2m}  \nonumber \\ \nonumber \\
\displaystyle\ \hspace{2cm} = -\mu\Omega_m^2\left(2-D(h)-D(-h)\right)^m,
\end{array}
\end{equation}
which has all good properties of a Laplacian (self-adjointness, ellipticity, negative semi-definiteness and
translational invariance).
The constitutive information due to the elastic potential (\ref{potenfu})
is contained in the ``Laplacian'' $\Delta_{2m}(h)$ which  has (by its definition the physical dimension of $[\mu\Omega_m^2]$, i.e. of a spring constant [force/length].
It is now straight forward to obtain the dispersion relation. Accounting for (\ref{shift}) yields for the generalized Laplacian (\ref{lapla}) (where $D(\frac{h}{2})-D(-\frac{h}{2})=2\sinh\frac{h}{2}\frac{d}{dx}$)

\begin{equation}
\label{lapla2}
\Delta_{2m}(h) = (-1)^{m+1} 4^m\mu\Omega_m^2 \sinh^{2m}\frac{h}{2}\frac{d}{dx}.
\end{equation}
Due to the periodic boundary conditions (\ref{periodic})
the $N$ vibrational eigenmodes are of the form of (ortho-normal) Bloch-vectors
\begin{equation}
\label{bloch}
 v_p(k_s) = \frac{e^{ik_sx_p}}{\sqrt{N}}= \frac{\eta^p}{\sqrt{N}},\hspace{0.5cm}\eta= e^{ik_sh} , \hspace{0.5cm}\ k_s=\frac{2\pi}{Nh}s,\, \,\,s=0,..,N-1,
\end{equation}
where $Nh=L$ denotes  the length of the linear chain and $x_p=ph$ the positions of the mass points. The Bloch vectors are equally eigenvectors of the (unitary) shift operator, namely $D(h)e^{ik_sx_p}= e^{ik_sh}e^{ik_sx_p}$ with the complex eigenvalues $e^{ik_sh}$ and realize irreducible representations of the cyclic group of shift operations of integer multiples of lattice spacing $h$. As any operator function of the shift operator such as (\ref{lapla}) has the same Bloch  eigenvectors, we obtain directly the dispersion relation $\omega_m^2(k_s)$ as (negative) eigenvalue of the generalized Laplacian, namely

\begin{equation}
\label{disperrel}
\Delta_{2m}(h)e^{ik_sx_p} = - \mu \omega_m^2(k_s) e^{ik_sx_p},
\end{equation}
which yields

\begin{equation}
\label{disper}
\omega_m^2(k_s) = \Omega_m^2 4^m\sin^{2m}{(\frac{k_sh}{2})} ,\hspace{1.5cm} k_s=\frac{2\pi}{Nh}s,\, \,\,s=0,..,N-1,
\end{equation}
where this model takes for $m=1$ the well known sine-square dispersion relation of the standard (next-neighbor) Born-von-Karman linear chain model.

Let us now consider general cases of non-local potentials of the form
(\ref{neighborsfunction}).
We then obtain by application of Hamilton's variational principle to this general potential its generalized ``Laplacian'' (note that
only contributions $m\geq 1$ are admissible), namely

\begin{equation}
\label{generalized}
\Delta_{f}(h)= -\mu \sum_{m=1}^{\infty}a_m\Omega_m^2\left(2-D(h)-D(-h)\right)^m = -\mu f\left(2-D(h)-D(-h)\right).
\end{equation}
In this relation we took into account the definition of ``characteristic function'' $f$ of (\ref{flam}).
The following picture may clarify how non-locality comes with (\ref{generalized}) into play: If we conceive
the displacement field as a $N$-component vector with Cartesian components $u_p$ (where $p=0,..,N-1$ indicates the mass point)
and let $h$ be the lattice spacing, then the shift operator $D(h)$ can be represented as matrix acting as follows

\begin{equation}
 \label{shiftgen}
 \begin{array}{l}
\displaystyle D(h)u_p=u_{p+1} =\sum_{q=0}^{N-1}\delta_{(p+1),q}u_q ,\hspace{0.5cm} D(-h)u_p=\sum_{q=0}^{N-1}\delta_{(p-1),q}u_q=u_{p-1}, \nonumber \\ \nonumber \\
\displaystyle p=0,..,N-1 \, ({\rm cyclically}),
\end{array}
\end{equation}
for the non-boundary points, and we {\it impose} the periodic boundary conditions (\ref{periodic}) also to the shift-operator

\begin{equation}
 \label{shift2}
D(h)u_{N-1}=u_{0} ,\hspace{1cm} D(-h)u_0=u_{N-1},
\end{equation}
Note that the ``Laplacian''
$\Delta_2= -\mu\Omega_2^2(2-D(h)-D(-h))$ of the Born-von-Karman chain corresponds to a ``Laplacian matrix'', namely
\begin{equation}
 \label{lapl2}
(\Delta_2(h))_{pq}=-\mu\Omega_2^2\left(2\delta_{pq}-\delta_{(p+1),q}-\delta_{(p-1),q}\right) ,\hspace{0.2cm} p=0,..,N-1 \,({\rm cyclically}),
\end{equation}
where its non-vanishing elements are ``localized'' around the diagonal. The equation 
(\ref{lapl2}) defines the entire Laplacian matrix of the next neighbor chain by assuming the cyclic index convention. 
We observe that (\ref{generalized}) constitute matrix functions
of the ``local" Laplacian matrix $(2-D(h)-D(-h)$ of (\ref{lapl2}). The matrix functions (\ref{generalized}) are defined by series of the integer powers (\ref{lapla})  of (\ref{lapl2}). These integer powers
represent matrix functions having non-zero elements being more delocalized from the diagonal the higher order $m$. They have the spectral representations
\begin{equation}
 \label{spectral}
 \begin{array}{l}
\displaystyle\ \Delta_{2m}(h) = -\mu \sum_{s=0}^{N-1}{\bf v}(k_s)\otimes {\bf v}^{*}(k_s) \omega_m^2(k_sh) \nonumber \\ \nonumber \\
\displaystyle\ \hspace{2cm} = -\mu\Omega_m^2\sum_{s=0}^{N-1}4^m\sin^{2m}{\frac{k_sh}{2}}\,\,{\bf v}(k_s)\otimes {\bf v}^{*}(k_s),
\end{array}
\end{equation}
where the $\omega_m^2(k_sh)$ denote the eigenfrequencies (\ref{disper}) and where we summarize over the entire set of $N$ Bloch-vectors (\ref{bloch}). $(*)$ indicates complex conjugation and $\otimes$ indicates dyadic multiplication. Note that all Laplacian matrices defined in the above way by Hamilton's principle are self-adjoint
and negative (semi) definite, translational invariant, and periodic (the matrices of the shift operators $D(\pm h)$ are the adjoint to each other).
The generally non-local Laplacian is generated by (\ref{generalized}) and yields a $N\times N$ matrix function which we can write in the spectral representation as

\begin{equation}
\label{lapmat}
\begin{array}{l}
\displaystyle \Delta_{f}(h)u(x_p) = -\frac{\partial }{\partial u_p}V_f ,\nonumber \\  \nonumber \\
\displaystyle \Delta_{f}(h)=-\mu f(2-D(h)-D(-h)) = -\mu \sum_{s=0}^{N-1}{\bf v}(k_s)\otimes {\bf v}^{*}(k_s)\omega_f^2(k_sh),
\end{array}
\end{equation}
with $\omega_f^2(k_sh)= f\left(4\sin^2(\frac{k_sh}{2})\right)$ where $f$ denotes the characteristic functions defined in (\ref{flam}). We can write for the Laplacian $\Delta_{f}(h)$ which takes the $N\times N$ Laplacian matrix representation
with the components
\begin{equation}
\label{constitmat}
\begin{array}{l}
\displaystyle\ (\Delta_{f}(h))_{pq} = (\Delta_{f}(h))(|p-q|) =
 -\frac{\mu}{N}\sum_{s=0}^{N-1} f\left(4\sin^2(\frac{k_sh}{2})\right)\,  e^{ik_sh(p-q)} ,\nonumber \\  \nonumber \\
\displaystyle\  k_s=\frac{2\pi}{Nh} s,
\end{array}
\end{equation}
depending only on $|p-q|$ which is a consequence of translational invariance. In general the entire Laplacian matrix (\ref{lapmat}) contains non-zero matrix elements which indicates non-locality.
The dispersion relation $\omega_f^2(k_sh)$ is defined by the series of the characteristic function (\ref{flam}), namely

\begin{equation}
\label{disprelaf}
\begin{array}{l}
\displaystyle \omega_f^2(k_sh) = f\left(4\sin^2(\frac{k_sh}{2})\right) =\sum_{m=1}^{\infty}a_m\Omega_m^2\left\{4\sin^2(\frac{k_sh}{2})\right\}^m \geq 0,   \nonumber \\ \nonumber \\
\displaystyle\ k_s=\frac{2\pi}{Nh}s ,\hspace{0.3cm} s=0,..,N-1,\nonumber \\ \nonumber \\
\displaystyle \omega_f^2(k_sh) =\omega_f^2(\kappa_s)  ,\hspace{1cm} \kappa_s=k_sh=\frac{2\pi}{N}s ,\hspace{0.3cm} s=0,..,N-1.
\end{array}
\end{equation}
The dispersion relation is explicitly only a function of the (non-dimensional) wave number
\begin{equation}
\label{kvar}
\kappa_s=k_sh=\frac{2\pi}{N}s ,\hspace{1cm} s=0,..,N-1.
\end{equation}
The non-dimensional wave number becomes (quasi-)continuous with values $0\leq \kappa_s\rightarrow \kappa <2\pi$ for $N>>1$ ''large".
The dispersion relation (\ref{disprelaf}) is positive $\omega_f^2(k_sh) >0$  for $k_s\neq 0$
and vanishing $\omega_f^2(k_sh=0) =0$ for $k_s=0$ reflecting translational invariance. This limits the admissible characteristic functions
$f$ to only characteristic functions $f(\lambda) >0$ ($0<\lambda <4$) and $f(\lambda=0)=0$ are admissible in order to generate physically admissible Laplacians $f(2-D(h)-D(-h))$.

So the elastic potential of the model (\ref{neighborsfunction}) is positive (semi-)definite and corresponds to the constitutive behavior of an elastically stable and translational invariant linear chain.
The spectral representation of the generalized Laplacian (\ref{lapmat}) yields the constitutive law in the form

\begin{equation}
 \label{constitutivelaw}
\Delta_{f}(h) \cdot {\bf u} = -\mu \sum_{s=0}^{N-1}{\bf v}(k_s)\omega_f^2(k_sh)({\bf u}\cdot{\bf v}(k_s)).
\end{equation}

The total elastic potential energy (\ref{potfunctionalf})
can be expressed by the Laplacian matrix as bilinear form of in the displacements ${\bf u}$

\begin{equation}
\label{poten}
\displaystyle V_f[u] = - \frac{1}{2}{\bf u}(\Delta_f(h)){\bf u} =
-\frac{1}{2} \sum_{p=0}^{N-1}\sum_{q=0}^{N-1}(\Delta_{f}(h))(|p-q|)u_pu_q \geq 0,
\end{equation}
where equality holds for uniform translations $u_p=const$ which do not contribute to the elastic energy reflecting translational invariance (corresponding to
$\omega_f^2(\kappa_s=0)=0$ for $\kappa_s=0$).
The Laplacian matrix (\ref{constitmat}) is a $N\times N$ symmetric and real valued (self-adjoint and negative-semi definite) matrix which depends only on the interparticle distance $|x_p-x_q|=h|p-q|$:

\begin{equation}
(\Delta_{f}(h))_{pq}=[\Delta_{f}(h)]_{qp}=\Delta_f(h|p-q|)=\Delta(|x_p-x_q|) ,\hspace{0.2cm} \Delta_f(h|p-q+N|),
\end{equation}
and is, due to the periodicity of the Bloch eigenvectors, also $N$-periodic with respect to both indices $p,q$ (or equivalently
$L$-periodic with respect to its dependence on $x=ph$).

Before we focus on continuum limits, let us now analyze the Laplacian matrix (\ref{constitmat}) for the above discrete linear chain in the limiting case $N\rightarrow \infty$ mass points, regardless to the interparticle spacing $h$.
The spectral representation of the Laplacian matrix can then be written as (\ref{constitmat}). The components of the Bloch vectors
(\ref{bloch}) depend only on the non-dimensional wave number $\kappa_s$ (\ref{kvar}) and can be written as

\begin{equation}
 \label{bloch2}
v_p(k_s)=[{\bf v}(\kappa_s)]_p=\frac{\eta^p(\kappa_s)}{\sqrt{N}}= \frac{e^{i\kappa_sp}}{\sqrt{N}} ,\hspace{0.5cm} \eta(\kappa_s) =e^{i\kappa_s} \hspace{0.2cm}, p=0,..,N-1.
\end{equation}

Thus the spectral representation of the $N\times N$-Laplacian matrix (\ref{lapmat}) can then be rewritten as

\begin{equation}
\label{lapmat2}
\begin{array}{l}
\displaystyle \Delta_{f} = -\mu \sum_{s=0}^{N-1}{\bf v}(\kappa_s)\otimes {\bf v}^{*}(\kappa_s)\omega_f^2(\kappa_s),\nonumber \\ \nonumber \\
\displaystyle \Delta_{f}(|p-q|)=-\frac{\mu}{N} \sum_{s=0}^{N-1} f(4\sin^2{\frac{\kappa_s}{2}}) e^{i\kappa_s(p-q)} ,\hspace{1cm} \kappa_s=\frac{2\pi}{N}s,
\end{array}
\end{equation}
 depending only on the set of non-dimensional wave numbers $\kappa_s$.

Let us now especially focus on the limiting case of large particle numbers $N>>1$:
A sum over the non-dimensional wave-numbers $\kappa_s$ can be rewritten by the asymptotic integral representation

\begin{equation}
 \label{asy}
\frac{1}{N}\sum_{s=0}^{N-1} g(\kappa_s) \approx \frac{1}{2\pi}\int_0^{2\pi}g(\kappa){\rm d}\kappa,
\end{equation}
where ${\rm d}\kappa\sim \kappa_{s+1}-\kappa_s = \frac{2\pi}{N}$ and $0\leq \kappa_s\rightarrow \kappa <2\pi$.
The Laplacian matrix (\ref{constitmat}) can then be rewritten as (with
$4\sin^2{\frac{\kappa}{2}}=2-e^{i\kappa}-e^{-i\kappa}$)

\begin{equation}
\label{laplacianmat}
\Delta_{f}(h)]_{pq} =\Delta_{f}(|p-q|) = -\frac{\mu}{2\pi}\int_0^{2\pi}f(2-e^{i\kappa}-e^{-i\kappa})e^{i\kappa(p-q)}{\rm d}\kappa.
\end{equation}

By introducing the complex variable $\xi=e^{i\kappa}$ and $0\leq \kappa< 2\pi$ this integral can be written as integral over the complex unit-circle $|\xi|=1$, and with $ie^{i\kappa}{\rm d}\kappa ={\rm d}\xi$

\begin{equation}
\label{laplacianmatcomplex}
\Delta_{f}(|p-q|) = -\frac{\mu}{2\pi i}\oint_{|\xi|=1}f(2-\xi-\xi^{-1}) \xi^{p-q-1}{\rm d}\xi.
\end{equation}

The complex integral representation (\ref{laplacianmatcomplex}) is especially convenient to determine the explicit representation of the Laplacian matrix for a ``large'' particle numbers $N>>1$ for a prescribed characteristic function $f(\lambda)$.
(\ref{laplacianmatcomplex}) recovers for $f(\lambda)= \Omega_2^2\lambda$ the Laplacian matrix of the Born-von-Karman linear chain with next neighbor springs:
Taking into account the Cauchy's residuum theorem

\begin{equation}
 \label{account}
\oint_{|\xi|=1} \xi^{n-1}{\rm d}\xi =2\pi i \delta_{n0} ,\hspace{1cm} n\in \ \mathbb{Z}_{0},
\end{equation}
being non-zero only for $n=0$, we obtain for $f(\lambda)= \Omega_2^2\,\lambda$ the Laplacian matrix
\begin{equation}
\label{laplacianmatcas2}
\begin{array}{l}
\displaystyle\ \Delta_{2}(|p-q|) = -\frac{\mu\Omega_2^2}{2\pi i}\oint_{|\xi|=1}(2-\xi-\xi^{-1}) \xi^{p-q-1}{\rm d}\xi  \nonumber \\ \nonumber \\
\displaystyle 
\displaystyle\ \hspace{2cm} = -\mu\Omega_2^2\left(2\delta_{pq}-\delta_{(p+1),q}-\delta_{(p-1),q}\right),
\end{array}
\end{equation}
in accordance with (\ref{lapla}) and its matrix representation (\ref{lapl2}) of the Born-von-Karman linear chain for $m=1$.

\subsection{Continuum limits of the Laplacian matrix}
\label{continuum}
In this section we consider two possible continuum limits where the Laplacian matrices take the forms of (non-local) convolution kernels. Throughout this analysis we denote with ``$\approx$" as equality in the limiting case $h\rightarrow 0+$ ($h>0$ remains infinitesimally non-zero positive):
\subsubsection{Periodic string continuum limit}
$h\rightarrow 0+$ ($h>0$ infinitesimal), $N(h)\sim h^{-1}\rightarrow \infty$ where the length of the string $L=N(h)h=const$ of the chain is kept finite and the $L$-periodicity of the chain is maintained.
\subsubsection{ The infinite medium continuum limit} This limiting case is obtained from (i) where the length of the chain tends to infinity $L\rightarrow \infty$.
\newline In both (quasi-)continuum limits (2.1.1) and (2.1.2) the discrete the coordinate $x_p$ of a particle $p$ becomes continuous where the particles appear as homogeneously distributed ''material points" with continuous coordinates $x$.

\subsection{(i) Periodic string continuum limit}
The first observation is that the dimensionless wave number is quasi-continuous (as $N=\frac{L}{h}\rightarrow \infty$) and hence relations
(\ref{asy})ff hold for summations in the (quasi-continuous) non-dimensional ($\kappa_s-$) wave number space whereas summations in the
space of {\it dimensional $k_s=\frac{\kappa_s}{h}=\frac{2\pi}{L}s$-wave numbers} remain discrete.
The spatial variable $0\leq x_p=hp < (N-1)h \approx L$  becomes quasi-continuous and sums over $x_p$
take the asymptotic forms of integrals over the quasi-continuous variable $x_p\rightarrow x$ and
$x_{p+1}-x_p =h\rightarrow {\rm d}x$, namely (where $\approx$ indicates asymptotic equality)
\begin{equation}
\label{summassp}
\sum_{p=0}^{N-1} g(x_p) \approx \frac{1}{h}\int_0^Lg(x){\rm d}x.
\end{equation}
We are interested in the continuum limit of bilinear forms such as the general potential energy (\ref{poten}). To this end we consider
\begin{equation}
\label{matlim}
\sum_{p=0}^{N-1}\sum_{q=0}^{N-1} G(x_p,x_q) \approx \frac{1}{h^2}\int_0^L\int_0^L G(x,x') {\rm d}x{\rm d}x' =
\int_0^L\int_0^L {\tilde G}(x,x'){\rm d}x{\rm d}x'.
\end{equation}
The discrete $N\times N$ matrix $G(x_p,x_q)$ corresponds to the continuum limit kernel

\begin{equation}
\label{kerneldef}
{\tilde G}(x,x') =  \lim_{h\rightarrow 0+} \frac{1}{h^2}G(x_p,x_q).
\end{equation}
Especially important is the localized kernel which correspond to $h\delta_{pq}$ which is obtained from
\begin{equation}
\label{unitdelta}
\sum_{p=0}^{N-1}\sum_{q=0}^{N-1}h\delta_{pq} =Nh=L =\int_0^L\int_0^L \delta(x-x'){\rm d}x{\rm d}x'.
\end{equation}
So we can link the unity matrix $(\delta_{pq})$ and the Dirac's $\delta$-function by the limiting case ($x_p\rightarrow x$,  $x_q\rightarrow x'$)
\begin{equation}
\label{limitingdelta}
\lim_{h\rightarrow 0+} \frac{1}{h^2}(h\delta_{pq})= \lim_{h\rightarrow 0+}\frac{1}{h}(\delta_{pq})=\delta(x-x').
\end{equation}

Employing relation (\ref{summassp}) on the particle number density (\ref{distfu}) becomes
\begin{equation}
\label{particledencont}
\rho_c(\xi)= \frac{1}{h}\int_0^L\delta(\xi-x){\rm d}x =\frac{1}{h}.
\end{equation}
and so indeed ${\tilde G}(x-x')\approx \rho_c(x)\rho_c(x')G(x_p,x_q)$. From the general relation (\ref{matlim}) follows for the continuum limit kernel which corresponds to the Laplacian matrix

\begin{equation}
\label{lalim}
{\tilde \Delta_f}(|x-x'|) = \lim_{h\rightarrow 0+} \frac{1}{h^2}\Delta_f(|x_p-x_q|).
\end{equation}
where $\Delta_f(|x_p-x_q|)=(\Delta_f(h))(|p-q|)$ denotes the components of the discrete Laplacian matrix (\ref{lapmat}).
We call in the following the corresponding continuum kernel ${\tilde \Delta_f}(|x-x'|)$ the ''{\it Laplacian kernel}".
Like the Laplacian matrix, the Laplacian kernel is translational invariant with respect to uniform translations of the string (corresponding to zero eigenfrequency of the $k=0$ mode) and hence depends only of the distance $|x-x'|$ of the material points.
The goal is now to determine general representations for the Laplacian kernel for prescribed characteristic functions $f$ (\ref{flam}). To this end consider the total elastic energy which can be written with (\ref{potfunctionalf})ff
in its quasi-continuous form

\begin{equation}
 \label{elasttotal}
\begin{array}{l} 
\displaystyle\ V_f=\frac{\mu}{2} \sum_{m=1}^{\infty}a_m\Omega_m^2\sum_{p=0}^{N-1}[(D(h)-1)^mu(x_p)]^2 \nonumber \\ \nonumber \\
\displaystyle\ \approx \frac{\mu}{2h} \sum_{m=1}^{\infty}a_m\Omega_m^2\int_0^L[(D(h)-1)^mu(x)]^2{\rm d}x,
\end{array}
\end{equation}
which is a bilinear functional of the (quasi-)continuous field $u(x)$ and where we applied here (\ref{summassp}) as the summation is performed over one index $p$~ only.
The asymptotic behavior of the $m$th order difference operator is given by
\begin{equation}
 (D(h)-1)^mu(x) =(e^{h\frac{d}{dx}}-1)^mu(x) \rightarrow h^{m}\frac{d^m}{dx^m}u(x) + 0(h^{m+1}).
\end{equation}
The terms $\sim h^{2m}\Omega^m$ are of the form
\begin{equation}
 \label{elasttotallimit}
V_f=\frac{\mu(h)}{2h} \sum_{m=1}^{\infty}a_m\Omega_m^2(h)h^{2m}\int_0^L\left(\frac{d^m}{dx^m}u(x)\right)^2{\rm d}x,
\end{equation}
remaining finite only if $\Omega_m^2(h)h^{2m}$ and $\frac{\mu(h)}{h}$ remain {\it finite} for $h\rightarrow 0$.
The finiteness of the continuum limit of a term of the order $m$ in the series (\ref{elasttotallimit}) requires
the following scaling behavior of the coefficients $\Omega_m^2(h)$ for $h\rightarrow 0$ (consistent with the scaling assumed in the local ($m=1$ type) chain model (Michelitsch et al., 2009):

\begin{equation}
\label{scaling}
\mu(h)=\rho_0 h, \hspace{1cm}
\Omega_m^2= A_mh^{-2m},
\end{equation}
where $\rho_0$ and $A_m$ are material constants independent of $h$. The first equation (\ref{scaling})$_1$ indicates a homogeneous distribution of mass where we introduced the constant mass density $\rho_0>0$.
In (\ref{scaling})$_2$ only the asymptotically order $\sim h^{-2m}$ is written (Orders $h^{-2n}$ $n<m$ lead to vanishing contributions and are therefore not relevant.). If a coefficient $\Omega^m(h)$ scales
as $\sim h^{-2m}$, it follows that all powers higher than $h^{2m}$ in $[(D(h)-1)^mu(x)]^2$ do not contribute, namely
\begin{equation}
\label{quasico}
\begin{array}{l} 
\displaystyle\ \lim_{h\rightarrow 0}\Omega_m^2(h)\left\{(D(h)-1)^mu(x)\right\}^2 = \lim_{h\rightarrow 0} A_m\left(\frac{(D(h)-1)^mu(x)}{h^m}\right)^2 \nonumber \\ \nonumber \\
\displaystyle\ \hspace{3cm} = A_m \left(\frac{d^m}{dx^m}u(x)\right)^2.
\end{array}
\end{equation}
There is no restriction on
integer $m$ in this relation (\ref{quasico}). In principle any {\it positive} $m>0 \in \mathbb{R}
$ including non-integers are admissible\footnote{As for any positive $m>0 \in\ R$ the wave-number zero mode has no elastic energy and hence translational invariance is preserved.}. For fractional $m \notin  \mathbb {N}$ the limiting case (\ref{quasico}) represents by evaluating the infinite fractional binomial series in the braces the definition of a {\it fractional derivative of order $m$} (which occurs symmetrized with respect to $h$) where the fractional operator acts in the limiting case as non-local convolutional kernel
on the field $u(x)$. The fractional case is, despite being an important source of non-locality, not subject of the present paper. We will devote to the analysis of fractional cases a sequel paper.
\newline\newline
The positive (renormalized material) constants $A_m>0$ do not scale with $h$ and can be defined by introducing two dimensional constants $\Lambda_m$ and $l_m$ by
\begin{equation}
 \label{charactlen}
A_m= \lim_{h\rightarrow 0+}h^{2m}\Omega_m^2(h) =\Lambda_m^2l_m^{2m}>0,
\end{equation}
where the $\Lambda_m$ and $l_m$ have the dimensions of frequencies and length-scales, respectively.
However we emphasize that
only the quantities $A_m$ is characteristic, whereas one of the quantities either frequency $\Lambda_m$ or length $l_m$ can be independently chosen such that $\Lambda_m^2l_m^{2m}=A_m$.
Depending on the scaling behavior of the material constants $\Omega_m^2(h)$ we can define the ``degree of non-locality" as follows:

We define as ``{\it weak non-locality}" if there is an order ${\hat m} \in \mathbb{N}$ such that $\Omega_m^2(h) \sim h^{-2m}$ scaling
as (\ref{scaling}) for $m\leq {\hat m}$, where all higher orders than $n > {\hat m}$ lead to vanishing limits $\Omega_n^2(h)h^{2n}\rightarrow 0$. The continuum limit of the Laplacian matrix in the case of weak
non-locality is then a polynomial of finite degree ${\hat m}$ in the local 1D standard Laplacian $\frac{d^2}{dx^2}$.
This case of weak non-locality leads to admissible models of gradient type such as by Lazar et al. (2006) and Maugin (1979).

In contrast we define as ``{\it strong non-locality}" when there are arbitrarily high orders $m$ where scaling relation (\ref{scaling}) is holding, so that the quasi-continuous limit yields an infinite series over $m$ (\ref{quasico}).
\newline\newline
Let us introduce characteristic coefficients taking either value $b_m=1$ if $h^{2m}\Omega_m^2(h)\approx A_m= \Lambda_m^2l_m^{2m}>0$ scale as (\ref{scaling})$_2$ or taking value $b_m=0$ if $h^{2m}\Omega_m^2(h)\rightarrow 0$. Diverging cases $h^{2m}\Omega_m^2(h)\rightarrow \infty$ where the continuum limit is not finite are not admitted.
With the renormalized material constants $A_m$ and the coefficients $b_m$, the total elastic energy of the string is obtained as

\begin{equation}
 \label{elasttorenorm}
V_f=\frac{\rho_0}{2} \int_0^L\left\{\sum_{m=1}^{\infty}a_mb_mA_m\left(\frac{d^m}{dx^m}u(x)\right)^2\right\}{\rm d}x,
\end{equation}

An important observation is the following: If $b_m=1$ for all orders $m$, ( i.e. if all coefficient $\Omega_m^2$ scale as (\ref{scaling}), then (\ref{elasttorenorm}) contains still the entire information
on the discrete lattice potential (and discrete Laplacian matrix). This property is extremely useful to reconstruct the full discrete lattice potential from continuum (long wave limit) dispersion relation data.
A typical example will be analyzed at the end of this paper.

From relation (\ref{elasttorenorm}) we will now determine the Laplacian kernel by reconsidering (\ref{poten}) which can be written in the quasi-continuous limit as

\begin{equation}
\label{poten1a}
\begin{array}{l}
\displaystyle V_f = - \frac{1}{2}{\bf u}(\Delta_f(h)){\bf u} = -\frac{1}{2} \sum_{p=0}^{N-1}\sum_{q=0}^{N-1}(\Delta_{f}(h))(|p-q|)u_pu_q ,\nonumber \\ \nonumber \\
\displaystyle V_f \approx  -\frac{1}{2}\int_0^L\int_0^L{\tilde \Delta_f}(|x-x'|)u(x)u(x'){\rm d}x{\rm d}x',
\end{array}
\end{equation}
where ${\tilde \Delta_f}(|x-x'|)$ denotes the {\it Laplacian kernel} which is defined by (\ref{lalim}) and ``$\approx$'' indicates asymptotic equality for $h\rightarrow 0$. We can write by using (\ref{lapmat}) together with (\ref{scaling}) the relation

\begin{equation}
\label{laplareno}
\begin{array}{l}
\displaystyle \int_0^L\int_0^Lu(x){\tilde \Delta_f}(|x-x'|)u(x'){\rm d}x{\rm d}x' \nonumber \\ \nonumber \\ \displaystyle =-\frac{\mu}{h}\int_0^Lu(x)\left(\sum_{m=1}^{\infty}a_mb_mA_m(-1)^mh^{-2m} (2\sinh\frac{h}{2}\frac{d}{dx})^{2m}\right)u(x){\rm d}x \nonumber \nonumber \\
\nonumber \\
\displaystyle =-\rho_0\int_0^L\int_0^Lu(x)
\left(\sum_{m=1}^{\infty}a_mb_mA_m(-1)^m\frac{d^{2m}}{dx^{2m}}\delta(x-x')\right)u(x'){\rm d}x{\rm d}x'.
\end{array}
\end{equation}

Comparison of the first and the last relation in (\ref{laplareno}) yields for the Laplacian kernel
(\ref{lalim})

\begin{equation}
\label{laplkerneldefinitiv}
{\tilde \Delta_f}(|x-x'|) =-\rho_0 \sum_{m=1}^{\infty}a_mb_mA_m(-1)^m\frac{d^{2m}}{dx^{2m}}\delta(x-x') = -\rho_0 {\tilde f}(-\frac{d^2}{dx^2})\delta(x-x'),
\end{equation}
which is defined as a convolutional kernel ``under the integral'' and
where ${\tilde f}$ indicates the ``{\it truncated characteristic function}"
\begin{equation}
\label{charact2}
\begin{array}{l}
 \displaystyle{\tilde f}(\lambda)= \lim_{h\rightarrow 0+} f_h(\lambda h^2) = \lim_{h\rightarrow 0+}\sum_{m=1}^{\infty}a_m \Omega_m^2(h)(h^2\lambda)^m, \nonumber \\ \nonumber \\
\displaystyle {\tilde f}(\lambda) =\sum_{m=1}^{\infty}a_mb_m A_m\lambda^m.
\end{array}
\end{equation}
The truncation appears if there are some coefficients $b_m=0$, indicating that there are some orders $m$ where $\Omega_m^2h^{2m}\rightarrow 0$.
It is important that elastic stability requirers that only such truncations are admissible such that ${\tilde f}(\lambda) \geq 0$ is maintained where equality holds for $\lambda=0$, especially elastic stability requires that ${\tilde f}(\lambda) >0$ {\it for all $\lambda >0$.} The equation 
(\ref{laplkerneldefinitiv}) can also be obtained by Hamilton's variation principle (functional derivative) with respect to field $u(x)$.
It is further important that all $\delta$-functions of the periodic string continuum limit (i) are $L$-periodic, i.e.  $L$-periodically continued, namely
\begin{equation}
 \label{deltaper}
\delta(x) =: \delta_L(x)= \delta_L(x+L),
\end{equation}
defining the unity operator in the function space of $L$-periodic functions.

Let us now also analyze the limiting case $h\rightarrow 0$ of the spectral representation of the Laplacian matrix
where we take into account the scaling assumption (\ref{scaling}). In this limit only fields $u(x_p\rightarrow x)$ are admissible which are ``smooth" compared to the interparticle spacing $h$, in order to guarantee (infinite) differentiability\footnote{such that the shift operator has representation (\ref{shift})} of $u(x)$ (Therefore continuum limits are equivalently referred to as the {\it long-wave limits}). That means
the lowest characteristic wave-length ${\tilde \lambda}_{min}(h) >>h$ associated with field $u(x)$
is much larger as the interparticle spacing.
For the following considerations it will be convenient when we choose the non-dimensional wave numbers within the first Brillouin zone $[-\pi,\pi]$
\begin{equation}
 \label{nondim}
-\pi \leq \kappa_s=\frac{2\pi}{N}s \leq \pi ,\hspace{0.5cm} s=\pm 0,\pm 1,\pm 2,..,\pm \frac{N}{2} \rightarrow \pm \infty \,\,(N>>1),
\end{equation}
and correspondingly the dimensional wave number is within {\it the (infinitely extended) first Brillouin zone $\frac{-\pi}{h}\rightarrow -\infty < k_s=\frac{\kappa_s}{h}=\frac{2\pi}{L}s < \frac{\pi}{h}\rightarrow \infty$}. In the long-wave limit not the entire set of $s$ of the first Brillouin zone is admitted.
To sufficiently smooth long-wave fields $u(x)$ only Bloch waves contribute
with wave numbers $\kappa_s$ where $|s|<|s_{max}| << N$ with wave-lengths  $\lambda_s =\frac{hN}{|s|}> \frac{hN}{s_{max}} = \lambda_{min} >> h$.
The long-wave spectral decomposition of the field writes then for $h<<1$ ``small''                   
\begin{equation}
\label{ured}
\begin{array}{l}
\displaystyle u_p =\sum_{s=-s_{max}(h)}^{s_{max}(h)}{\hat u}(\kappa_s)\frac{1}{\sqrt{N}}e^{i\kappa_sp} ,\hspace{0.5cm} \frac{s_{max}(h\rightarrow 0+)}{N} << 1 , \hspace{0.5cm} (s_{max}(h) >>1),\nonumber \\ \nonumber \\
\displaystyle {\bf u}={\cal P}(h)\cdot{\bf u}, \nonumber \\ \nonumber \\
\displaystyle {\cal P}(h)=\sum_{s=-s_{max}(h)}^{s_{max}(h)<<N} {\bf v}(\kappa_s)\otimes {\bf v}^*(\kappa_s),
\end{array}
\end{equation}
where ${\cal P}(h)$ denotes the unity operator (projection operator) in this reduced smooth ``long-wave function space". Applied on any field vector ${\bf u'}$, the projection operator retains only its long-wave part ${\bf u} = {\cal P}(h)\cdot{\bf u'}$
where (\ref{ured})$_2$ holds if ${\bf u'}={\bf u}$ contains only long wave modes.
In (\ref{ured}) we further introduced the modal amplitudes
${\hat u}(\kappa_s)={\bf u}\cdot{\bf v}^*(\kappa_s)$. The projection operator is an incomplete (long wave) part of the $N\times N$ unity operator $\delta_{pq}$. As the latter, the projection operator depends only on $|p-q|$ and is
$N$-periodic ${\cal P}(h)_{|p-q|} = {\cal P}(h)_{|p-q+N|}$.
The reduced wave space of the long wave limit guarantees
that $u$ is sufficiently smooth that the shift operator indeed can be represented by the Taylor series
$D(h)u(x)=e^{h\frac{d}{dx}}u(x)=u(x+h)$. It makes sense to analyze in the limiting case of $h\rightarrow 0$ the implicit $h$-dependence dependence of the projection operator ${\cal P}(h)$. For convenience we call the function space spanned by
${\cal P}$ as ``{\it long wave space}''. It is important that in the long wave space
$s_{max}(h) \rightarrow \infty$ tends to infinity, however ``weaker'' than $N(h)=Lh^{-1}\rightarrow \infty$, such that $\frac{s_{max}(h)}{N(h)} \rightarrow 0$. Without loss of generality we can generate this behavior by assuming the scaling

\begin{equation}
 \label{scalingsmax}
s_{max}(h) = -Const\, \log(h)  \sim \log(N) >>1 ,\hspace{1cm} 0<h<<1,
\end{equation}
with a $Const >0$ and independent on $h$. 
In order to determine the limit of the Laplacian matrix it is useful to analyze first
the dispersion relation (\ref{disprelaf}) for $\kappa_s << 1$ and by taking into account the scaling relations
(\ref{scaling}). Let us write therefore the characteristic function in the form $f(\lambda)=f(h,\lambda)$ where the dependence on $h$ indicates the implicit dependence of the coefficients $\Omega_m^2(h)$ on $h$ and where we have to consider the limiting case
$h\rightarrow 0+ (h\neq 0)$ in the long-wave space

\begin{equation}
\label{cocomp}
\begin{array}{l}
\displaystyle \omega_f^2(\kappa_s)=f(h,4\sin^2{\frac{\kappa_s}{2}}), \nonumber \\ \nonumber \\
\displaystyle \omega_f^2(\kappa_s) =\sum_{m=1}^{\infty}a_m\Omega_m^2(h)[2(1-\cos{\kappa_s})]^m=
\sum_{m=1}^{\infty}a_m\Omega_m^2(h)\kappa_s^{2m}(1-2\frac{\kappa_s^2}{4!}+..)^{m},\nonumber \\ \nonumber \\
\displaystyle \omega_f^2(\kappa_s)= \sum_{m=1}^{\infty}a_m A_m h^{-2m}(\kappa_s^{2m} +O(\kappa_s^{2m+2})), \nonumber \\ \nonumber \\ 
\displaystyle \kappa_s=\frac{2\pi}{N}s ,\hspace{1cm}  -s_{max}(h)\leq s\leq s_{max}(h) \sim \log(N) << N .

\end{array}
\end{equation}

Now it is important that in the long wave limit the {\it non-dimensional wave number  $-\kappa_{s_{max}}\leq \kappa_s\leq  \kappa_{s_{max}} \sim  -h\log(h) \rightarrow 0$   is in the limiting case within an infinitely small interval}

\begin{equation}
\label{Kasnon}
\begin{array}{l}
-\frac{2\pi}{N}s_{max}\leq \kappa_s \leq \kappa_{s_{max}}= \frac{2\pi}{N}s_{max} =\frac{2\pi}{L}hs_{max} \sim -h\log(h) << 1, \nonumber \\ \nonumber \\
-k_{smax}\leq k_s=\frac{\kappa_s}{h} \leq k_{smax} = \frac{2\pi}{L}s_{max} \sim s_{max}(h) \sim -\log(h) \sim \log(N) >> 1,
\end{array}
\end{equation}
whereas the {\it dimensional wave number $k_s$ can take any finite value $- k_{s_{max}} \leq k_s=\frac{\kappa_s}{h} \leq k_{s_{max}} \sim -\log(h) \rightarrow \infty$}.
We have therefore the cases

\begin{equation}
\label{cons}
\begin{array}{l}
\displaystyle \Omega_m^2(h)\kappa_s^{2m+2n} \approx A_m (\frac{\kappa_s}{h})^{2m} \kappa_s^{2n} \approx A_m k_s^{2m} \kappa_s^{2n}
=A_mk_s^{2m}\nonumber \\ \nonumber \\
\displaystyle= A_m(\frac{2\pi}{L}s)^{2m} \leq Const'\log^{2m}(h) ,\nonumber \\ \nonumber \\
\displaystyle\ {\rm finite \,\,\, for }  \hspace{0.2cm}n=0, and \nonumber \\ \nonumber \\
\displaystyle\Omega_m^2(h)\kappa_s^{2m+2n} \approx A_mk_s^{2m}\kappa_s^{2n} \leq Const''h^{2n} \log^{2m+2n}(h)\,\, \rightarrow 0  ,\nonumber \\ \nonumber \\
{\displaystyle \rm vanishing \,\,\, for} \hspace{0.2cm} n=1,2,.. >0.
\end{array}
\end{equation}
For $n=0$ the asymptotic values of (\ref{cons}) are finite for $s$ fixed where the upper limit becomes unbounded $0\leq \Omega_m^2(h)\kappa_s^{2m} \leq \Omega_m^2(h)\kappa_{s_{max}}^{2m} \sim \log^m(N)\rightarrow \infty$.
 It follows in the series
(\ref{cocomp}) only the terms $\kappa_s^{2m}\Omega_m$ contribute, namely

\begin{equation}
 \Omega_m^2(h)[2(1-\cos{\kappa_s})]^m \approx A_mk_s^{2m} ,\hspace{0.5cm} -\infty <k_s=\frac{\kappa_s}{h}=\frac{2\pi}{L}s <\infty.
\end{equation}

As a consequence the dispersion relation (\ref{cocomp}) takes the (renormalized) form of the long wave limit
\begin{equation}
 \label{renormdisp}
 \begin{array}{l}
\displaystyle{\tilde \omega}_f^2(k_s)= \lim_{h\rightarrow 0+} f(h,4\sin^2{\frac{k_sh}{2}}) = {\tilde f}(k_s^2)= \sum_{m=1}^{\infty}b_ma_mA_mk_s^{2m},\nonumber \\ \nonumber \\
\displaystyle k_s=\frac{2\pi}{L}s ,\hspace{0.5cm} -\infty <s<\infty,
\end{array}
\end{equation}
with the {\it truncated characteristic function ${\tilde f}$ of (\ref{charact2})}.
Applying the continuum limit (\ref{charact2}) we notice that application of ${\tilde f}(-\frac{d^2}{dx^2})$ on the Bloch eigenfunctions indeed leads to

\begin{equation}
{\tilde \Delta}_f e^{ik_sx} =-\rho_0{\tilde f}(k_s^2)e^{ik_sx} = -\rho_0 {\tilde \omega}_f^2(k_s),
\end{equation}
with the renormalized long-wave limit dispersion relation (\ref{renormdisp}).

In the {\it non-truncated case} (${\tilde f}=f$ the scaling relation (\ref{scaling})$_2$ holds for all non-zero coefficients $\Omega_m^2$ thus all coefficients $b_m=1$) the full constitutive information of the characteristic function is in the long wave limit preserved and in principle observable. Especially the elastic potential
(\ref{potfunctionalf})
of the corresponding discrete chain model can be reconstructed from determining the (coefficients of the) characteristic function by long-wave dispersion relation (inelastic Neutron scattering) measurements.
We will give an explicit benchmark example at the end of this paper.

We can now obtain the Laplacian kernel also in the long wave space by its spectral representation from the limit
(\ref{lalim})

\begin{equation}
\label{specl}
{\tilde \Delta_f}(x_p-x_q) = -\frac{\mu}{h^2} \sum_{s=-s_{max}(h)}^{s_{max}(h)} f(h,4\sin^2{\frac{\kappa_s}{2}})v_p(\kappa_s)v_q^*(\kappa_s).
\end{equation}
First we have $\mu=\rho_0h$ and let us take into account that
the renormalized Bloch vectors are obtained by
\begin{equation}
 \label{blochscal}
{\bf v}(k_s)\cdot{\bf v}^*(k_l) =\frac{1}{N}\sum_{p=0}^{N-1}e^{i(k_s-k_l)x_p}\approx \frac{1}{L}\int_0^Le^{\frac{2\pi i x}{L}(s-l)}{\rm d}x =\delta_{sl},
\end{equation}
where $k_s=\frac{2\pi s}{Nh}= \frac{2\pi s}{L}$ and $Nh=L$. The renormalized Bloch eigenmodes of the continuum limit are
\begin{equation}
 \label{renormbloch}
 \begin{array}{l}
\displaystyle\phi_s(x)=\frac{1}{\sqrt{L}}e^{ik_sx}= \frac{1}{\sqrt{h}}v_p(\kappa_s) ,\hspace{0.5cm} x_p=x,  \nonumber \\ \nonumber \\
 \displaystyle\ k_s=\frac{2\pi s}{L}=\frac{\kappa_s}{h} , (s=0,\pm 1,..,\pm s_{max}(h) \sim -\log(h) )\rightarrow \infty,
\end{array}
\end{equation}
so that $\phi_s(x)\phi^*_s(x') = \frac{1}{h}v_p(\kappa_s)v_q^*(\kappa_s)$ ($x=x_p$ and $x'=x_q$).
The Laplacian kernel takes then the form

\begin{equation}
\label{kernellap}
\begin{array}{l}
\displaystyle {\tilde \Delta_f}(x-x')  =-\rho_0 \sum_{s=-s_{max}(h)}^{s_{max}(h)}{\tilde f}(k_s^2)\phi_s(x)\phi^*_s(x'), \hspace{0.3cm} s_{max}(h) \sim -\log(h) \rightarrow \infty,\nonumber \\ \nonumber \\
\displaystyle {\tilde \Delta_f}(x-x')=-\frac{\rho_0}{L} \sum_{s=-\infty}^{\infty}{\tilde \omega}^2(k_s)e^{ik_s(x-x')} ,\hspace{1cm} k_s=\frac{2\pi}{L}s,
\end{array}
\end{equation}
and hence
\begin{equation}
\begin{array}{l}
\label{laplkerneldefinitivlimit}
\displaystyle {\tilde \Delta_f}(|x-x'|) =-\rho_0 \sum_{m=1}^{\infty}a_mb_mA_m(-1)^m\frac{d^{2m}}{dx^{2m}}\delta(x-x') \nonumber \\ \nonumber \\
\displaystyle\ = -\rho_0{\tilde f}(-\frac{d^2}{dx^2})\delta(x-x'),
\end{array}
\end{equation}
coinciding with (\ref{laplkerneldefinitiv}).
Note that the long wave space projection operator ${\cal P}(h)$ defined in (\ref{ured}) takes asymptotically the {\it complete} (L-periodic) $\delta$-function

\begin{equation}
\label{reno}
\begin{array}{l}
\displaystyle \lim_{h\rightarrow 0+}\frac{1}{h}{\cal P}(h) = \lim_{h\rightarrow 0+}\sum_{s=-smax(h)}^{s_{max}(h)}\frac{1}{h}{\bf v}(\kappa_s)\otimes{\bf v}^*(\kappa_s) \approx \delta(x-x'), \nonumber \\ \nonumber \\
\displaystyle \delta(x-x')=\delta_L(x-x')= \sum_{s=-\infty}^{\infty} \phi_s(x)\phi^*_s(x') =\frac{1}{L}\sum_{s=-\infty}^{\infty} e^{ik_s(x-x')},\nonumber \\ \nonumber \\
\displaystyle\ k_s=\frac{2\pi}{L}s,
\end{array}
\end{equation}
and application of $-\rho_0{\tilde f}(-\frac{d^2}{dx^2})$ on (\ref{reno}) recovers again (\ref{kernellap}).
Hence we finally arrive for the spectral representation of the {\it Laplacian kernel} at

\begin{equation}
\label{reladif}
\begin{array}{l}
\displaystyle {\tilde \Delta}_f(|x-x'|) = -\rho_0\sum_{m=1}^{\infty}b_ma_mA_m(-1)^m\frac{d^{2m}}{dx^{2m}}\delta(x-x')\nonumber \\ \nonumber \\
\displaystyle  = -\frac{\rho_0}{L}\sum_{s=-\infty}^{\infty}\sum_{m=1}^{\infty}b_ma_mA_m k_s^{2m} e^{ik_s(x-x')} \nonumber \\ \nonumber \\
\displaystyle = - \rho_0\sum_{s=-\infty}^{\infty} {\tilde \omega}_f^2(k_s)\phi_s(x)\phi^*_s(x').
\end{array}
\end{equation}

We notice that the Laplacian kernel is an {\it even function} in $x-x'$ reflecting spatial isotropy in 1D.

\subsection{(ii) Infinite medium continuum limit $L \rightarrow \infty$}
In this limit the relations for the Laplacian kernel (\ref{reladif}), (\ref{laplkerneldefinitivlimit}) remain
the same, only the boundary conditions change from $L$-periodic to those of the ``infinite medium''. The dimensional wave number $k_s$ in (\ref{reladif}) becomes quasi-continuous
taking values $k_s\rightarrow k$ with $0<k<\infty$ and the $\delta$-function becomes the $\delta$-function
of the infinite 1D space.
So a sum over $-\infty< k_s=\frac{2\pi}{L}s \rightarrow k <\infty$   with ${\rm d}k \sim k_{s+1}-k_s =\frac{2\pi}{L}$ can be asymptotically written as integral

\begin{equation}
\label{kasum}
\sum_{s=-\infty}^{\infty} g(k_s) \approx \frac{L}{2\pi}\int_{-\infty}^{\infty}g(k){\rm d}k,
\end{equation}

This relation can be applied to the $L$-periodic Bloch eigenfunctions

\begin{equation}
\begin{array}{l}
\displaystyle \sum_{s=-\infty}^{\infty} \phi_s(x)\phi^*_s(x') =\frac{1}{L}\sum_{s=0}^{\infty} e^{ik_s(x-x')} =\delta_L(x-x`) ,\hspace{0.3cm} k_s=\frac{2\pi}{L}s , \hspace{0.3cm} L\rightarrow \infty,
\nonumber \\ \nonumber \\
\displaystyle \approx \frac{1}{2\pi}\int_{-\infty}^{\infty} e^{ik(x-x')}{\rm d}k = \int_{-\infty}^{\infty} \Phi_k(x)\Phi_k^*(x'){\rm d}k= \delta(x-x'),
\end{array}
\end{equation}
which is a well known expression for the infinite space $\delta$-function which leads to the renormalized infinite space Bloch-eigenfunctions
\begin{equation}
 \Phi_k(x)=  \sqrt{\frac{L}{2\pi}}\phi_s(x)= \frac{e^{ikx}}{\sqrt{2\pi}} ,\hspace{1cm} -\infty < k < \infty.
\end{equation}

Applying (\ref{kasum}) to (\ref{reladif}) we can write for the Laplacian kernel of the infinite medium the spectral
representation

\begin{equation}
\label{reladifinfimed}
\begin{array}{l}
\displaystyle {\tilde \Delta}_f(|x-x'|) = -\rho_0\sum_{m=1}^{\infty}b_ma_mA_m(-1)^m\frac{d^{2m}}{dx^{2m}}\delta(x-x')  \nonumber \\ \nonumber \\
\displaystyle =-\frac{\rho_0}{2\pi}\int_{-\infty}^{\infty}\left(\sum_{m=1}^{\infty}a_mb_mA_m k^{2m}\right)e^{ik(x-x')}{\rm d}k \nonumber \\
\nonumber \\
\displaystyle = -\rho_0 \int_{-\infty}^{\infty} {\tilde \omega}_f^2(k)\Phi_k(x)\Phi_k^*(x'){\rm d}k,
\end{array}
\end{equation}
where (\ref{reladifinfimed})$_3$ is form invariant with (\ref{reladif})$_3$ in the renormalized Bloch eigenmodes $\phi_s\rightarrow \Phi_k$.
Let us briefly establish the link between the ``distributional representation" (\ref{reladifinfimed}) of the Laplacian kernel
(\ref{reladifinfimed}) and its explicit spatial representation. To this end we consider a constitutive convolution
\begin{equation}
\label{constitutiveconvol}
\int_{-\infty}^{\infty}{\tilde \Delta}_f(|x-x'|)u(x'){\rm d}x' = \int_{-\infty}^{\infty}{\tilde \Delta}_f(|\xi|)\frac{u(x+\xi)+u(x-\xi)}{2}{\rm d}\xi,
\end{equation}
where we have used that the Laplacian kernel is an even function
which we rewrite by means of the shift operator as
\begin{equation}
\label{convolshift}
\begin{array}{l}
\displaystyle\frac{1}{2}(u(x+\xi)+u(x-\xi)) = \frac{1}{2}(D_x(\xi)+D_x(-\xi))u(x) \nonumber \\ \nonumber \\
\displaystyle\ =\cosh{(\xi\frac{d}{dx})}\,\,u(x)=\sum_{m=0}^{\infty}\frac{\xi^{2m}}{(2m)!}\frac{d^{2m}}{dx^{2m}}u(x),
\end{array}
\end{equation}
with this relation we can write for convolution (\ref{constitutiveconvol})
\begin{equation}
\label{constitutiveconvol2}
\int_{-\infty}^{\infty}{\rm d}\xi{\tilde \Delta}_f(|\xi|)\cosh{(\xi\frac{d}{dx})}   u(x)=
\sum_{m=0}^{\infty}\left(\int_{-\infty}^{\infty}{\tilde \Delta}_f(|\xi|)\frac{\xi^{2m}}{(2m)!}{\rm d}\xi\right)\frac{d^{2m}}{dx^{2m}}u(x),
\end{equation}
 where the term of $m=0$ is vanishing\footnote{reflecting translational invariance, i.e. the $k=0$ Bloch wave function has no elastic energy.} due to

\begin{equation}
\int_{-\infty}^{\infty}{\tilde \Delta}_f(|\xi|){\rm d}\xi =0.
\end{equation}

It is now straight-forward to relate the moment integrals in the braces of (\ref{constitutiveconvol2}) with the coefficients of the (truncated) characteristic function ${\tilde f}$ by writing
(\ref{constitutiveconvol2}) in the identical form
\begin{equation}
\label{identfo}
\int_{-\infty}^{\infty}{\rm d}x'u(x')\left[\left(\int_{-\infty}^{\infty}{\rm d}\xi{\tilde \Delta}_f(|\xi|)\cosh{(\xi\frac{d}{dx})}\right)\delta(x-x')\right].
\end{equation}
By comparison with (\ref{constitutiveconvol}) the expression in the braces $[..]$ can be identified again with the Laplacian kernel
(\ref{reladifinfimed}).
Plugging in the $\delta$-function in its spectral representation and by using that $\cosh{\xi\frac{d}{dx}}e^{ikx}=\cos{k\xi}e^{ikx}$ we obtain
\begin{equation}
\label{powerser}
-\rho_0{\tilde f}(k^2)= \int_{-\infty}^{\infty}{\tilde \Delta}_f(|\xi|)\cos{(k\xi)}\,{\rm d}\xi.
\end{equation}
Comparing the orders in $k^{2m}$ yields
\begin{equation}
\label{termsinm}
(-1)^{m+1}\rho_0A_ma_mb_m= \int_{-\infty}^{\infty}{\tilde \Delta}_f(|\xi|)\frac{\xi^{2m}}{(2m)!}{\rm d}\xi ,\hspace{1cm} m=1,2,.. \mathbb{N},
\end{equation}
i.e. this relation for the moment integrals coming into play in (\ref{constitutiveconvol2})
 with the coefficients of the characteristic function again is consistent with representation (\ref{reladifinfimed}) for the Laplacian kernel. Note that the order $m=0$ of (\ref{termsinm}) is vanishing.
The moments of integrals (\ref{termsinm}) give to the coefficient of the (truncated) characteristic function in a sense a ``physical" interpretation.

\noindent {\it  (a) Non-local Hooke's law}\newline
 Now it is only straight-forward to establish from the negative semi-definite Laplacian kernel (\ref{reladifinfimed}), the positive definite elastic modulus kernel 
$C_f(|x-x'|)$ when we take into account that the elastic energy can be represented as a bilinear functional
in the deformations $\frac{du(x)}{dx}$, namely

\begin{equation}
\label{elasten}
V_f = \frac{1}{2}\int_{-\infty}^{\infty} \int_{-\infty}^{\infty}C_f(|x-x'|)\frac{du(x)}{dx}\frac{du(x')}{dx'}{\rm d}x{\rm d}x'.
\end{equation}
By taking into account (\ref{renormdisp})  for the infinite medium, $L\rightarrow \infty$
we have

\begin{equation}
\label{elastenfou}
\begin{array}{l}
\displaystyle \rho_0{\tilde \omega}_f^2(k)= k^2{\tilde C}_f(k) =  k^2\rho_0\sum_{m=1}^{\infty}a_mb_mA_mk^{2m-2}, \nonumber \\ \nonumber \\
\displaystyle {\tilde C}_f(k)= \rho_0\sum_{m=1}^{\infty}a_mb_mA_mk^{2m-2},
\end{array}
\end{equation}
where the elastic modulus kernel $C_f(|x-x'|)$ is then given by
\begin{equation}
\label{elastmodker}
\begin{array}{l}
\displaystyle C_f(|x-x'|) = -\rho_0\sum_{m=1}^{\infty}b_ma_mA_m(-1)^m\frac{d^{2m-2}}{dx^{2m-2}}\delta(x-x') = \nonumber \\ \nonumber \\
\displaystyle  =\hspace{0.3cm}\rho_0 b_1a_1A_1 \delta(x-x')- \rho_0 b_2a_2A_2\frac{d^{2}}{dx^{2}}\delta(x-x')+..\nonumber \\ \nonumber \\
\displaystyle\ +\rho_0 b_ma_mA_m(-1)^{m-1}\frac{d^{2m-2}}{dx^{2m-2}}\delta(x-x')+..
\hspace{0.5cm}.
\end{array}
\end{equation}

 The elastic modulus kernel and Laplacian kernel are connected by the simple relation $\frac{d^2}{dx^2}C_f(|x-x'|)= {\tilde \Delta}_f(|x-x'|)$. In (\ref{elastmodker}) the first term for $m=1$ which is due to the local Born-von-Karman next neighbor chain corresponds to local linear (standard) elasticity with the elastic modulus $C_1$. If this contribution is nonzero in our model we have $a_1=b_1=1$ and so the standard elastic modulus is given by
\begin{equation}
\label{elaststandard}
C_1=\rho_0 A_1=\rho_0\Omega_2^2 h^2 > 0.
\end{equation}

\subsection{Reconstruction of the lattice model from long-wave dispersion relation data}

Let us assume the non-truncated case ${\tilde f}=f$ where the full constitutive information of the characteristic function is preserved in the long wave dispersion relation.
The long-wave limit dispersion relation, (i.e. in principle the constants $A_m$) are experimentally accessible from inelastic Neutron scattering measurements. Let us assume that all nonzero constants scale $\Omega_m^2\approx \frac{A_m}{h^{2m}}$ and further $k\approx \frac{\kappa_s}{h}$ as well as the particle mass $\mu=\rho_0h$ and particle number $N$ and lattice constant $h$ are known.
Then we can reconstruct from the long wave dispersion relation the full dispersion relation of the entire
Brillouin zone by (and in this way we can reconstruct the entire discrete lattice elastic potential)

\begin{equation}
\label{reconstruct}
\begin{array}{l}
\displaystyle\ \hspace{1cm} (a)\hspace{0.3cm} long \,\, wave \,\, limit \,\,of \,\, discrete \,\,chain:\nonumber \\ \nonumber \\
\displaystyle {\tilde \omega_f^2}(k)= \sum_{m=1}^{\infty}a_mA_mk^{2m} \approx \sum_{m=1}^{\infty}a_m\frac{A_m}{h^{2m}}\kappa_s^{2m} =f(\kappa_s^2)\nonumber \\ \nonumber \\
\displaystyle\ \hspace{1cm} (b) \hspace{0.3cm} entire \,\, Brillouin-zone\,\, s=0,\pm 1,\pm 2,..\pm \frac{N}{2}:\nonumber \\ \nonumber \\
\displaystyle f(\kappa_s^2) \rightarrow f(4\sin^2{\frac{\kappa_s}{2}}) = \sum_{m=1}^{\infty}a_m\frac{A_m}{h^{2m}} 4^m\sin^{2m}{\frac{\kappa_s}{2}} =\omega_f^2(\kappa_s)\nonumber \\ \nonumber \\

\end{array}
\end{equation}
 i.e. from the long-wave data $A_m$ we can reconstruct the {\it entire dispersion relation (characteristic function)  $\omega_f^2(\kappa_s)$ (\ref{disprelaf}) of the complete first Brillouin zone (\ref{reconstruct})$_2$}
and hence also the corresponding elastic potential which corresponds to that discrete chain, with the potential which corresponds to (\ref{potfunctionalf}) (where $\mu=\rho_0h$), namely

\begin{equation}
 \label{reconpot}
V_f= \frac{\mu}{4} \sum_{p=0}^{N-1}\sum_{m=1}^{\infty} a_m\frac{A_m}{h^{2m}}\left\{\left[(D(h)-1)^mu(x_p)\right]^2+\left[(D(-h)-1)^mu(x_p)\right]^2\right\}.
\end{equation}

\subsubsection{Example}
An important non-local elastic modulus kernel which is often introduced phenomenologically
is the Gaussian kernel, (see e.g.~in the book of Eringen (2002) and others). We assume there is no truncation of the characteristic function in the continuum limit.
Let us hence solve the inverse problem. We determine here the entire dispersion relation of the first Brillouin zone and the elastic potential, respectively,
of the discrete chain which corresponds to the long wave data due of a Gaussian elastic modulus kernel. The Gaussian kernel is assumed in the form
\begin{equation}
 C_g(|x-x'|) = \frac{C_0}{2\pi}\int_{-\infty}^{\infty}e^{ik(x-x')}e^{-ak^2}{\rm d}k = C_0 \frac{e^{-\frac{(x-x')^2}{4a}}}{\sqrt{4\pi a}},
\end{equation}
with constants $a>0$, $C_0>0$. This Gaussian kernel is  normalized
\begin{equation}
\int_{-\infty}^{\infty}C_g(|x-x'|){\rm d}x' =C_0,
\end{equation}
where the Fourier transformed kernel ${\tilde C}_g(k)$ which is defined in the same way as (\ref{elastenfou})
is given by
\begin{equation}
 \label{gaussfou}
{\tilde C}_g(k) = C_0e^{-ak^2}= C_0\sum_{n=0}^{\infty}\frac{(-a)^n}{n!}k^{2n} = \frac{\rho_0\omega_g^2(k)}{k^2},
\end{equation}
with the long wave dispersion relation and characteristic function $f_g$
\begin{equation}
 \label{omegafg}
{\tilde \omega}_g^2(k)= \frac{k^2}{\rho_0}{\tilde C}_g(k) =  \frac{C_0}{\rho_0}k^2e^{-ak^2} \approx \frac{C_0}{\rho_0}\frac{\kappa_s^2}{h^2}e^{-a\frac{\kappa_s^2}{h^2}} =f_g(\kappa_s^2)
\end{equation}
and hence
\begin{equation}
\label{coefficientsAm}
 a_mA_m= (-1)^{m-1}\frac{C_0}{\rho_0} \frac{a^{m-1}}{(m-1)!} ,\hspace{2cm} m=1,2,..,\ N,
\end{equation}
where $a_m=(-1)^{m-1}$ denote the signs of the coefficients.
The dispersion relation of the entire Brillouin zone
is then reconstructed by replacing  $\kappa_s^2 \rightarrow 4\sin^2{\frac{\kappa_s}{2}}$ in (\ref{coefficientsAm})
and yields
\begin{equation}
 \label{entire}
 \begin{array}{l}

\displaystyle  \omega_g^2(\kappa_s) = f_g(4\sin^2{\frac{\kappa_s}{2}}) =\omega_0^2 \sin^2{\frac{\kappa_s}{2}} \, e^{(-4\gamma\sin^2{\frac{\kappa_s}{2}})}
\nonumber \\ \nonumber \\
\displaystyle  \omega_0^2=\frac{4C_0}{h^2\rho_0} ,\hspace{0.2cm} \gamma=\frac{a}{h^2} >0,\hspace{0.2cm} \kappa_s=\frac{2\pi}{N}s ,\hspace{0.2cm} s=0\pm 1, \pm 2,..,\pm \frac{N}{2}.
\end{array}
\end{equation}
In this reconstructed complete dispersion relation relation we introduced the dimensionless parameter $\gamma=\frac{a}{h^2} >0 $ which is a measure for nonlocality and the dimensionless wave numbers $\kappa_s$ take (for $N>>1$ large quasi-continous) values
within the first Brioullin zone $-\pi \leq \kappa_s=\kappa \leq \pi$. We therefore skip in the following the subscript and put $\kappa_s\rightarrow \kappa$.
The elastic potential $V_g$ of relation (\ref{potfunctionalf}) which corresponds is the given by series (\ref{reconpot}) with the coefficients (\ref{coefficientsAm})
and writes

\begin{equation}
 \label{reconpotgauss}
  \begin{array}{l}
\displaystyle V_g= \frac{hC_0}{4} \sum_{p=0}^{N-1}\sum_{m=1}^{\infty}\frac{(-1)^{m-1}}{(m-1)!}\frac{a^{m-1}}{h^{2m}}\left\{\left[(D(h)-1)^mu(x_p)\right]^2\right. \nonumber \\ \nonumber \\
\displaystyle +\left[(D(-h)-1)^mu(x_p)\right]^2 \}.
\end{array}
\end{equation}
Then
the dispersion relation (106) yields the group speed

\begin{equation}
 \label{vitesse}
v_g(\kappa_s)=  h\frac{d\omega_g}{d \kappa_s} =v_0 \cos{\frac{\kappa_s}{2}}\,
\left(1-4\gamma\sin^2{\frac{\kappa_s}{2}}\right)\, e^{-2\gamma\sin^2{\frac{\kappa_s}{2}}} . \hspace{0.5cm} v_0=\sqrt{C_0/\rho_0} 
\end{equation}
Let us determine the zeros within the first Brillouin zone: We have at the boundary for any admissible characteristic function (\ref{flam}) always vanishing group speed at the boundaries of the first Brioullin zone

\begin{equation}
 \label{zerosboundary}
  \cos{\frac{\kappa}{2}} = 0 ,\hspace{1cm} \kappa=\pm \pi,
\end{equation}
The vanishing of the group speed at the boundaries of the first Brillouin zone is a necessary ``good property'' imposed by
the periodicity of the chain.
The second case of vanishing group speeds
\begin{equation}
\label{and}
1-4\gamma\sin^2{\frac{\kappa}{2}} = 0 , \hspace{1cm}   \sin^2{\frac{\kappa}{2}} = \frac{1}{4\gamma}   \leq 1,
\end{equation}
exists only for $\gamma=\frac{a}{h^2} \geq \frac{1}{4}=\gamma_t$ and yields

\begin{equation}
 \label{zeros}
      \kappa(\gamma) = \pm 2\arcsin{\frac{1}{2\sqrt{\gamma}}}.
\end{equation}
The two extreme cases exist $\kappa(\gamma_t=\frac{1}{4})= \pm \pi$ and in the limiting case of ``extreme nonlocality'' $\gamma >> 1$ we asymptotically
$\kappa(\gamma >>1) \approx \pm \frac{1}{\sqrt{\gamma}}\rightarrow 0$.

For $\gamma \geq \gamma_t=\frac{1}{4}$ the dispersion relation (\ref{entire})
has maxima occuring symetrically at the $\kappa$-values (\ref{zeros})
with the maximum values

\begin{equation}
 \label{disprelextrema}
\omega_g^2(\kappa(\gamma))= \frac{v_0^2}{a e} =\frac{v_0^2}{\gamma h^2 e} \sim \frac{1}{\gamma}
\end{equation}
where $e$ denotes the Euler number and scaling as $\sim \frac{1}{\gamma}$ when $h$ is kept fixed.

The appearence of the maxima of the dispersion relation occuing symetrically at at (\ref{zeros})
above the critical value of $\gamma=\frac{a}{h^2}\geq \frac{1}{4}$ is a characteristic feature of a non-local constitutive law of gaussian type. In the critical case these maxima are at the boundary of the Brillouin zone at 
$\kappa(\gamma=\frac{1}{4})= \pm \pi$ where the $|\kappa|$-values of these maxima (\ref{zeros}) 
are descreasing monotoneously as the non-locality parameter $\gamma$ is increasing. 
For ``extreme nonlocality'' $\gamma >> 1$ the posititions of these maxima approach asymptotically 
$\kappa(\gamma >>1) \approx \pm \frac{1}{\sqrt{\gamma}}\rightarrow 0$. The case of extreme nonlocality is discussed in below case II.
\\ \\
The following two limiting cases are important:
\\ \\
\noindent {\bf Case I: Local limit: $h$ finite and $a\rightarrow 0+$, $\gamma=\frac{a}{h^2}<<1$, }: \\

Then the elastic kernel becomes localized
\begin{equation}
 C(|x-x'|)\approx C_0\delta(x-x'),
\end{equation}
and takes the form of a Dirac's $\delta$-function
recovering conventional local elasticity and the dispersion relation (106) takes asymptotically the form ($A_1=v_0^2$)

\begin{equation}
 \label{limit}
\omega_g^2(\kappa)= \omega_0^2\sin^2(\frac{\kappa}{2}) 
\end{equation}
coinciding with the dispersion relation of the local Born-von-Karman chain.
\\ \\
\noindent {\bf Case II: Extreme nonlocal limit:  $h$ finite, $a>>1$ thus $\gamma >>1 $}: \\
The Gaussian modulus function becomes extremely delocalized and the dispersion extremely localized
with asymptically quadractic behavior $\omega_g^2(\kappa_s) \sim \kappa_s^2$ within $-\frac{1}{\sqrt{\gamma}} <\kappa_s< \frac{1}{\sqrt{\gamma}}$ where $\pm \frac{1}{\sqrt{\gamma}}$ are the asymptotic $\kappa_s$-values of the maxima
and where the dispersion relation decays rapidly to zero outside this interval with maxima at
\begin{equation}
\label{broadlimit}
\omega_g^2(\gamma=\frac{a}{h^2} >>1) \approx \omega_g^2(\kappa_s=\pm \frac{1}{\sqrt{\gamma}}) \approx 
\frac{v_0^2}{h^2}\kappa_s^2e^{-\gamma\kappa^2}|_{\kappa=\frac{1}{\sqrt{\gamma}}} = \frac{v_0^2}{ae} ,
\end{equation}
where $e=e^{\xi}|_{\xi=1}$ denotes the Euler number. The elastic modulus then is
so delocalized that the microstructure (lattice spacing $h$) becomes invisible.
The only nonzero eigenfrequencies in this limit $\gamma=\frac{a}{h^2}>>1$ are $\omega_g=\pm \sqrt{{v_0^2}/{\ ae}}$.


 
Lazar et al. (2006) analyzed the dispersion relation which is obtained from Eringen's nonlocal elasticity model (Eringen, 1983) of the type bi-Helmholtz type. By matching the bi-Helmholtz dispersion relation with the Born-von-Karman dispersion relation, they found that a bi-Helmholtz model with two parameters is required to match best a physically admissible dispersion relation.

The advantage of our approach is that always physically admissible dispersion relations are obtained with
vanishing group velocity (zeros (\ref{zerosboundary})) at the boundaries of the Brillouin zone for any admissible characteristic function (\ref{flam}) and any positive nonlocality parameter $\gamma$ (see Fig. 2) without any matching to other lattice models.
Furthermore, for values of the nonlocality parameter greater than the transition value $\gamma_t = \frac{1}{4}$, the dispersion spectrum (see Fig. 1) presents a softening similar to those which one observes experimentally in the case of longitudinal phonons in direction $[110]$ of metallic cubic (fcc) crystals and insulators  (see for instance Eringen 1972, Bliz et al 1979).

\section{Conclusions}
We analyzed $N$-periodic linear chains in 1D with non-local harmonic interactions introduced by elastic potentials
defined by quadratic forms of differences of order $m=1,2,..\in \ \mathbb{N}$ of the particle displacements which
include $m$ neighbor particles (symmetrically in $\pm$-directions). From those basic elastic potentials we generated series which include in general all orders of $m$ and generate in this way arbitrary non-locality of the harmonic interactions and constitutive laws.
We focused on series of integer orders $m\in \ \mathbb{N}$ (where order $m=0$ is vanishing due to our assumption of translational invariance). There is no restriction to differences of integer orders $m$: Any positive fractional $m>0 \in \ \mathbb{R}$ are also
admissible, where the even order derivatives (integer-powers of the standard Laplacian) are then replaced by fractional powers of the Laplacian (Some models of non-local constitutive behavior of {\it fractional type} and their interconnection to self-similar chain models were established recently (Michelitsch et al., 2009a, 2009b,2012; Michelitsch, 2011).
Fractional order models of this type will be analyzed in a sequel paper.

Application of Hamilton's variational principle on the above mentioned elastic potentials
defines generally non-local ''Laplacian operators`` which are constituted by even order-differences of orders $2m$ of the field. In the $N$-dimensional vector-space of
particle displacements $(u_p)={\bf u}$, these generalized Laplacians act as $N\times N$ matrices and constitute negative (semi-) definite self-adjoint matrix functions
of the ``nearby'' diagonal local Laplacian matrix of the next neighbor (Born-von-Karman linear chain model) represented by order $m=1$. Increasing orders of $m$ in the elastic potentials produce in the displacement field vector space ``increasingly'' non-diagonal $N\times N$ {\it Laplacian matrices} being the source of non-local constitutive behavior.

We analyze series generally including all orders $m$ which involve in the displacement space highly non-diagonal Laplacian matrices leading to general non-local constitutive behavior. The analysis includes continuum limits such as the periodic string limit (i) and the infinite medium continuum limit (ii), respectively, and allows to define different degrees of non-locality: week non-locality when the continuum limit includes only finite orders $m$ and strong non-locality when an infinite sequence of orders $m$ ``survives" the continuum limit. The weak case leads to models of ``strain gradient elasticity''. In contrast the cases of strong non-locality allow to represent arbitrary non-local constitutive behavior. To establish a finite continuum limit of the elastic energy, the assumption of scaling relations (\ref{scaling})$_2$ which renormalize the material constants are crucial. This includes already the lowest order $m=1$ of next neighbor Born-von-Karman linear chain model which yields in the continuum limit the 1D standard Laplacian operator with local standard elasticity.

The present approach has potential to be extended in several directions: It is to be extended to fractional cases of (positive) non-integer $m$ in 1D and it can be extended to multi-dimensional, especially periodic (Bravais) lattices. In general our approach can be extended to all cases where the eigenmodes of local problems are known, for instance for 2D and 3D lattice dynamics models which take into account only close neighbor springs (Askar, 1985; Maradudin et. al. 1963). The corresponding discrete Laplacians of such local models can then be ``delocalized" by our approach.

We hope that the present paper inspires further
analysis of non-locality, especially as to account for nonlocality appears to have a vast potential of new physical- and nano-scale engineering applications.
\newline\newline
\noindent {\bf Acknowledgements.} The authors are indepted to G\'erard Maugin (University Pierre and Marie Curie) as well as to No\"el Challamel (University of South Brittany) and CW Lim (City University of Hong Kong) for stimulating discussions.
\\

\noindent {\bf References}\\
\\
\noindent Askar, A. (1982). A generalization of the Korteweg- de-Vries equation for anharmonic lattices and vectorial solitons, International  Journal of Engineering Science, 20, 169-179.\\
\\
\noindent Askar A. (1985) {\it Lattice dynamical foundations of continuum theories, elasticity, piezoelectricity, viscoelasticity, plasticity}. Singapore, World Scientific.\\
\\
\noindent Bilz H. \& Kress W. (1979). {\it Phonon dispersion relations in insulators}, Solid State Science 10, Berlin, Springer-Verlag.\\
\\
\noindent Born M., Huang K. (1954). {\it Dynamical theory of crystal lattices. London, Oxford University Press}. \\
\\
\noindent Cadet S. (1987). Coupled transverse-longitudinal enveloppe modes in an atomic chain, Journal of physics C: Solid State Physics, 20, L803-L811. \\
\\
\noindent   Challamel N., Wang C.M. (2008). The small length scale effect for a non-local cantilever beam: a
paradox solved, Nanotechnology 19, 345703.\\
\\
  \noindent Collet B. (1993). Lattice Approach of Shear Horizontal Solitons in Cubic Cristal Plates. Material Science Forum Vol. 123-125 , 417-426.\\
\\
 \noindent Eringen A.C. (1972). Linear theory of nonlocal elasticity and dispersion of plane waves. Int. J. Eng.
   Sci. 10, 425--435 .\\
\\
\noindent Eringen A.C., Kim B.S. (1977). Relation between Non-Local Elasticity and Lattice Dynamics. Crystal Lattice Defects 1977, 7, 51-57.\\
\\
 \noindent Eringen A.C. (1983). On differential equations of nonlocal elasticity and solutions of screw dislocation and
    surface waves. J. Appl. Phys. 54(9), 4703-4710.\\
\\
 \noindent Eringen A.C. (1992).  Vistas of nonlocal continuum physics. International Journal of Engineering Science, 30, 1551-1565.\\
\\
\noindent Eringen A.C. (2002). {\it Nonlocal continuum field theories}. New York: Springer-Verlag.\\
\\
\noindent Hu, Y. G., Liew, K.M., Wang, Q. , He, X. Q. \& Yakobson, B. I., (2008). Nonlocal shell model for elastic wave propagation in single-and double-walled carbon nanotubes, Journal of the Mechanics and Physics of Solids, 56, 3475-3485.\\
\\
\noindent Kr\"oner E. (1967). Elasticity theory of materials with long range cohesive forces. International Journal of Solids and Structures 3, 731-742.\\
\\
\noindent Krumhansl J.A. (1968). {\it Some Considerations of the Relation between Solid State Physics and Generalized Continuum Mechanics}. In: Mechanics of Generalized Continua, Proceedings of the IUTAM-Symposium, Freudenstadt and Stuttgart (Germany) 1967, E. Kr\"oner Ed., Springer, Berlin, Heidelberg, New York.\\
\\
\noindent  Kunin I.A. (1982). {\it Elastic media with microstructure I and II}. New York, Springer-Verlag. \\
\\
 \noindent  Lazar M., Maugin G.A., Aifantis E.C. (2006). Theory of nonlocal elasticity of bi-Helmholtz type and some applications. International. Journal of Solids and Structures 43, 1404-1421.\\
\\
\noindent Lim C. W. (2010). On the truth  of nanoscale for nanobeams based on nonlocal elastic stress field theory: equilibrium, governing equation and static deflection. Applied  Mathematics and Mechanics (English Edition), 31, 37-54.\\  
\\
\noindent Lu, P., Zhang, P.Q., Lee, H.P., Wang, C.M., Reddy, J.N., (2007). Non-local elastic plate theories. Proceeding of Royal. Society . A 463, 3225-3240. \\
\\
\noindent  Maugin G.A. (1979). Nonlocal theories or gradient type theories: a matter of convenience? Archives Mechanics 31, 15-26.\\
\\
\noindent Maugin, G. A. (1999). {\it Nonlinear waves in elastic crystals}, Oxford, Oxford University Press. \\
\\
\noindent Maradudin A.A., Montroll E.A., Weiss G.N. (1963). {\it Theory of lattice dynamics in the harmonic approximation}. Solid State of Physics. , Academic Press New York.\\
\\
\noindent Michelitsch T.M., Maugin G.A., Nowakowski A.F., Nicolleau F.C.G.A. (2009a).
Analysis of the vibrational spectrum of a linear chain with spatially exponential properties. International Journal of Engineering Science 4, 209-220.\\
\\
\noindent Michelitsch T.M., Maugin G.A., Nicolleau F.C.G.A., Nowakowski A.F., Derogar S. (2009b).
Dispersion Relations and Wave Operators in Self-Similar Quasi-Continuous Linear Chains. Physical
Review E 80, 011135.\\
\\
\noindent Michelitsch T.M., Maugin G.A., Rahman M., Derogar S., Nowakowski A.F., Nicolleau F.C.G.A. (2012). An approach to generalized one-dimensional self-similar elasticity. Int. J. Eng. Sci. 61, 103-111.\\
\\
\noindent Michelitsch T.M., Maugin G.A., Rahman M., Derogar S., Nowakowsky A.F., Nicolleau F.C.G.A. (2012). A continuum theory for one-dimensional self-similar elasticity and applications to wave propagation and diffusion, European Journal of Applied Mathematics 23(12), 709-735.\\
\\   
\noindent Michelitsch T.M. (2011). The self-similar field and its application to a diffusion problem.
J. Phys. A: Math. Theor. 44, 465206.\\
\\
\noindent Pouget J. (2005). Non-linear lattice models: complex dynamics, pattern formation and aspects of chaos, Philosophical Magazine 85, Numbers 33-35, 4067-4094.\\
\\
\noindent Reddy J. N. (2007) Nonlocal theories for bending, buckling and vibration of beams, International Journal of Engineering Science 45, 288-307.\\
\\
\noindent Remoissenet M. \& Flytzanis N. (1984). Solitons in anharmonic chains with long range interactions, Journal of physics C: Solid State Physics, 18, 85, 4603-4629.\\
\\
\noindent Rosenau P (2003). Hamiltonian dynamics of dense chains and lattices: or how to correct the continuum. Physics Letters A 311, 39-52.\\
\\
\noindent Scrivastava, G. P. (1990).{\it The physics of phonons},  Bristol, Adam Hilger (IOP).
\\
\noindent Zhang Y.Q., Liu R.G. , Xie X. Y. (2005). Free transverse vibrations of double-waled carbon nanotubes using a theory of non local elasticity. Physical Review B 71, 195404.

\end{document}